%% file: main.tex
\documentclass[fleqn,usenatbib,letters]{mnras}

\usepackage{newtxtext,newtxmath}
\usepackage[T1]{fontenc}
\usepackage{graphicx}
\usepackage{xcolor}
\usepackage{xspace}
\usepackage{amsmath}
\usepackage{booktabs}
\usepackage{tabularx}
\hypersetup{hypertexnames=false}

\DeclareRobustCommand{\VAN}[3]{#2}
\let\VANthebibliography\thebibliography
\def\thebibliography{\DeclareRobustCommand{\VAN}[3]{##3}\VANthebibliography}

\let\oldAA\AA
\renewcommand{\AA}{{\text{\normalfont\oldAA}}}
\makeatletter
\renewcommand{\ion}[2]{#1\,{\scshape\@roman{#2}}}
\makeatother

\newcommand{\lya}{Ly$\alpha$\xspace}
\newcommand{\ha}{H$\alpha$\xspace}
\newcommand{\hb}{H$\beta$\xspace}

\newcommand{\civ}{\ion{C}{4}\xspace}
\newcommand{\heii}{\ion{He}{2}\xspace}
\newcommand{\oii}{[\ion{O}{2}]\xspace}
\newcommand{\oiii}{[\ion{O}{3}]\xspace}

\newcommand{\piii}{Pop~{III}\xspace}
\newcommand{\kms}{\,\mathrm{km\,s^{-1}}}

\newcommand{\logoh}{12+\log(\mathrm{O/H})}
\newcommand{\Rthree}{R3\equiv[\mathrm{O\,III}]\,\lambda5007/\mathrm{H}\beta}
\newcommand{\RthreeResult}{$0.59^{+0.27}_{-0.21}$}
\newcommand{\zlya}{$z_{\mathrm{Ly}\alpha}=4.8001\pm0.0002$}
\newcommand{\zprism}{$z_{\rm PRISM}=4.800\pm0.003$}

\newcommand{\HeIIUVResult}{$F_\mathrm{He\,II\,\lambda1640, 2\sigma}<4.45\times10^{-19}\,\mathrm{erg\,s^{-1}\,cm^{-2}}$\xspace}
\newcommand{\HeIIOptResult}{$F_\mathrm{He\,II\,\lambda4686, 2\sigma}<1.33\times10^{-19}\,\mathrm{erg\,s^{-1}\,cm^{-2}}$\xspace}
\newcommand{\target}{LATED-1\xspace}

\input{author}

\title[JWST IFU of LATED-1]{LATED: JWST integral field spectroscopy of a galaxy caught in chemical infancy at $z=4.8$ behind Abell 2744}

\author[M. Li et al.]{
\parbox{\textwidth}{\raggedright
\orcidsymb{Mingyu Li}{0000-0001-6251-649X}\affil{THU,KICC,Cavendish}\corrmark,
\orcidsymb{Roberto Maiolino}{0000-0002-4985-3819}\affil{KICC,Cavendish,UCL},
\orcidsymb{Zheng Cai}{0000-0001-8467-6478}\affil{THU},
\orcidsymb{Hannah \"Ubler}{0000-0003-4891-0794}\affil{MPE},
\orcidsymb{Boyuan Liu}{0000-0002-4966-7450}\affil{Heidelberg},
\orcidsymb{Qiao Duan}{0009-0009-8105-4564}\affil{KICC,Cavendish},
\orcidsymb{Fuyan Bian}{0000-0002-1620-0897}\affil{ESO,CASSACA},
\orcidsymb{Sijia Cai}{0009-0003-4133-0292}\affil{THU},
\orcidsymb{Francesco D'Eugenio}{0000-0003-2388-8172}\affil{KICC,Cavendish},
\orcidsymb{Eiichi Egami}{0000-0003-1344-9475}\affil{Steward},
\orcidsymb{Bjorn H.~C.~Emonts}{0000-0003-2983-815X}\affil{NRAO},
\orcidsymb{Xiaohui Fan}{0000-0003-3310-0131}\affil{Steward},
\orcidsymb{Yuki Isobe}{0000-0001-7730-8634}\affil{KICC,Cavendish,Waseda},
\orcidsymb{Lucy R. Ivey}{0009-0002-5105-1222}\affil{KICC,Cavendish},
\orcidsymb{Xihan Ji}{0000-0002-1660-9502}\affil{KICC,Cavendish},
\orcidsymb{Gareth C. Jones}{0000-0002-0267-9024}\affil{KICC,Cavendish},
\orcidsymb{Maria Koller}{0009-0000-1950-9112}\affil{KICC,Cavendish},
\orcidsymb{Xiaojing Lin}{0000-0001-6052-4234}\affil{THU},
\orcidsymb{Christopher C. Lovell}{0000-0001-7964-5933}\affil{KICC,IoA},
\orcidsymb{Kimihiko Nakajima}{0000-0003-2965-5070}\affil{Kanazawa},
\orcidsymb{Masami Ouchi}{0000-0002-1049-6658}\affil{NAOJ,ICRR,SOKENDAI,IPMU},
\orcidsymb{Robert G. Pascalau}{0000-0001-9820-5773}\affil{KICC,Cavendish},
\orcidsymb{J. Xavier Prochaska}{0000-0002-7738-6875}\affil{UCSC,IPMU},
\orcidsymb{Jan Scholtz}{0000-0001-6010-6809}\affil{KICC,Cavendish},
\orcidsymb{Fengwu Sun}{0000-0002-4622-6617}\affil{Westlake},
\orcidsymb{Sandro Tacchella}{0000-0002-8224-4505}\affil{KICC,Cavendish},
\orcidsymb{Yunjing Wu}{0000-0003-0111-8249}\affil{IPMU,CD3},
\orcidsymb{Fujiang Yu}{0000-0002-3489-6381}\affil{THU}, and 
\orcidsymb{Zijian Zhang}{0000-0002-2420-5022}\affil{KIAA,PKU,KICC,Cavendish}
}
\vspace{0.4cm}\\
\parbox{\textwidth}{Affiliations are listed at the end of the paper.\corremail{lmytime@hotmail.com}}
}

\defaffil{THU}{Department of Astronomy, Tsinghua University, Beijing 100084, People’s Republic of China}
\defaffil{KICC}{Kavli Institute for Cosmology, University of Cambridge, Madingley Road, Cambridge, CB3 0HA, UK}
\defaffil{Cavendish}{Cavendish Laboratory - Astrophysics Group, University of Cambridge, 19 JJ Thomson Avenue, Cambridge, CB3 0HE, UK}
\defaffil{UCL}{Department of Physics and Astronomy, University College London, Gower Street, London WC1E 6BT, UK}
\defaffil{ESO}{European Southern Observatory, Alonso de C\'ordova 3107, Casilla 19001, Vitacura, Santiago 19, Chile}
\defaffil{CASSACA}{Chinese Academy of Sciences South America Center for Astronomy, National Astronomical Observatories, CAS, Beijing 100101, People’s Republic of China}
\defaffil{Edinburgh}{Institute for Astronomy, University of Edinburgh, Royal Observatory, Blackford Hill, Edinburgh EH9 3HJ, UK}
\defaffil{NOIRLab}{NSF's National Optical-Infrared Astronomy Research Laboratory, 950 N. Cherry Avenue, Tucson, AZ 85719, USA}
\defaffil{Steward}{Steward Observatory, University of Arizona, 933 North Cherry Avenue, Tucson, AZ 85721, USA}
\defaffil{NRAO}{National Radio Astronomy Observatory, 520 Edgemont Road, Charlottesville, VA 22903, USA}
\defaffil{Waseda}{Waseda Research Institute for Science and Engineering, Faculty of Science and Engineering, Waseda University, 3-4-1, Okubo, Shinjuku, Tokyo 169-8555, Japan}
\defaffil{Heidelberg}{Universit\"at Heidelberg, Zentrum f\"ur Astronomie, Institut f\"ur Theoretische Astrophysik, Albert-Ueberle-Str. 2, D-69120 Heidelberg, Germany}
\defaffil{IoA}{Institute of Astronomy, University of Cambridge, Madingley Road, Cambridge, CB3 0HA, UK}
\defaffil{Kanazawa}{Institute of Liberal Arts and Science, Kanazawa University, Kakuma-machi, Kanazawa, Ishikawa 920-1192, Japan}
\defaffil{NAOJ}{National Astronomical Observatory of Japan, 2-21-1 Osawa, Mitaka, Tokyo 181-8588, Japan}
\defaffil{ICRR}{Institute for Cosmic Ray Research, The University of Tokyo, 5-1-5 Kashiwanoha, Kashiwa, Chiba 277-8582, Japan}
\defaffil{SOKENDAI}{Department of Astronomical Science, SOKENDAI (The Graduate University for Advanced Studies), 2-21-1 Osawa, Mitaka, Tokyo 181-8588, Japan}
\defaffil{IPMU}{Kavli Institute for the Physics and Mathematics of the Universe (WPI), The University of Tokyo Institutes for Advanced Study, The University of Tokyo, Kashiwa, Chiba 277-8583, Japan}
\defaffil{UCSC}{Department of Astronomy and Astrophysics, University of California, Santa Cruz, 1156 High Street, Santa Cruz, CA 95064, USA}
\defaffil{Westlake}{Department of Astronomy, School of Science, Westlake University, Hangzhou, Zhejiang 310030, People’s Republic of China}
\defaffil{CfA}{Center for Astrophysics $|$ Harvard \& Smithsonian, 60 Garden St., Cambridge, MA 02138, USA}
\defaffil{MPE}{Max-Planck-Institut f\"ur extraterrestrische Physik (MPE), Gie{\ss}enbachstra{\ss}e 1, 85748 Garching, Germany}
\defaffil{CD3}{Center for Data-Driven Discovery, Kavli IPMU (WPI), UTIAS, The University of Tokyo, Kashiwa, Chiba 277-8583, Japan}
\defaffil{KIAA}{Kavli Institute for Astronomy and Astrophysics, Peking University, Beijing 100871, People’s Republic of China}
\defaffil{PKU}{Department of Astronomy, School of Physics, Peking University, Beijing 100871, People’s Republic of China}

\date{}
\pubyear{2026}

\begin{document}
\label{firstpage}
\pagerange{\pageref{firstpage}--\pageref{lastpage}}
\maketitle

\begin{abstract}
When Population~III (Pop III) star formation ended remains an open question.
\target is an intrinsically faint ($M_\mathrm{UV}=-16.15$) \lya emitter at $z=4.80$ revealed by VLT/MUSE behind the lensing cluster Abell~2744 ($z=0.308$).
Before any spectroscopic metallicity constraints, it was identified as an extremely metal-poor or metal-free galaxy candidate from JWST imaging by LATED, our novel photometric selection framework.
Here we present serendipitous JWST/NIRSpec PRISM integral-field spectroscopy of this target.
The spectrum reveals \lya, \hb, and \ha at 7.0, 5.6, and 17.7$\sigma$, as well as tentative detections of \oiii$\lambda\lambda4959,5007$ at 2.9$\sigma$, and yields $R3=\mathrm{[O\,III]\lambda5007/H\beta}=0.59^{+0.27}_{-0.21}$, upholding the earlier LATED photometric prediction of $R3<1.54$ ($2\sigma$ limit).
The canonical JWST-based strong-line calibration, extrapolated to low metallicity, implies $\logoh=6.45^{+0.17}_{-0.19}$, or $Z/Z_\odot=0.58_{-0.20}^{+0.28}\%$, placing \target among the most metal-poor galaxies known.
Its modest magnification ($\mu=2.80$) leaves the intrinsic properties insensitive to the lens model.
\target is therefore a dwarf galaxy caught in the earliest state of chemical enrichment, and a compelling target for testing whether Pop III star formation can persist to $z<5$.
The result demonstrates that photometric selection, especially through the LATED methodology, can reach below one per cent solar metallicity to identify Pop III galaxy candidates for spectroscopic follow-up.
\end{abstract}


\section{Introduction}
\label{sec:introduction}

Population~III (\piii) stars are the first generation of stars formed from primordial gas in the early Universe \citep{Bromm2004ARAA..42...79B,Bromm2013RPPh...76k2901B,Klessen2023ARAA..61...65K}.
As the first supernovae enriched their surroundings, the pristine gas able to form \piii stars became progressively rarer and more isolated, surviving only in pockets where metals had not yet mixed.
How long such pockets persist, and hence when \piii star formation effectively ceased, is not known.
Early models ended the \piii era by $z\sim15$ through widespread pre-enrichment \citep{Yoshida2004ApJ...605..579Y}, whereas later semi-analytic models and simulations let it continue past $z\sim6$ \citep{Mebane2018MNRAS.479.4544M} or until reionization ends at $z\simeq5$ \citep{Zier2025MNRAS.544..410Z}.
Others allow it to persist to lower redshift wherever metal mixing remains inefficient \citep{Scannapieco2003ApJ...589...35S, Tornatore2007MNRAS.382..945T, Maio2010MNRAS.407.1003M, Xu2016ApJ...823..140X, Liu2020MNRAS.497.2839L}.
Answering the question of when \piii star formation ceases observationally is also challenging: \piii signatures are short-lived, rapidly diluted by self-enrichment, and imitable by ordinary Population~II stars in very metal-poor gas \citep{Schaerer2003A&A...397..527S, Raiter2010A&A...523A..64R, Katz2023MNRAS.524..351K,Rusta2025ApJ...989L..32R}.
The most credible observational searches therefore combine a metallicity constraint ($Z/Z_\odot\lesssim1\%$) with an independent probe of the hard ionizing spectrum, particularly narrow high-equivalent-width \heii emission \citep{Feltre2016MNRAS.456.3354F, Plat2019MNRAS.490..978P, Nakajima2022MNRAS.513.5134N, Trussler2023MNRAS.525.5328T}.
Even under the conservative \piii criterion adopted here, no source satisfying both conditions has yet been identified so far.
Meanwhile, JWST spectroscopy has extended gas-phase oxygen abundances to $z\sim10$ and, through lensing, to stellar masses below $10^{7}\,M_\odot$ \citep{Nakajima2023ApJS..269...33N, Curti2024A&A...684A..75C, Sanders2024ApJ...962...24S, Chemerynska2024ApJ...976L..15C, Mowla2024Natur.636..332M,Ubler2026arXiv260320360U}.
Yet the measured abundances almost never fall below $\logoh\simeq7.0$ or $Z/Z_\odot\simeq2\%$.
The deepest direct-temperature stacks bottom out at two per cent solar, and $z>5$ samples cluster above an apparent floor at 1--2 per cent solar \citep{Isobe2026arXiv260611345I,Hsiao2025arXiv250503873H}.
The floor admits two scenarios: the first, supernovae may genuinely drive every surviving galaxy past one per cent solar within its first star-forming episodes, or surveys that select and confirm galaxies through their brightest metal lines may be unable to register anything beneath it.
The few systems reported below the metallicity floor cannot decide between these scenarios, because each was found through extreme lensing magnification or serendipity rather than selected systematically in advance.
The benchmarks with $Z/Z_\odot < 1\%$ are the $\mu\simeq100$ stellar complexes LAP1-B at $z=6.625$ \citep[$Z/Z_\odot=(0.42\pm0.18)\%$,][Scholtz et al., in preparation]{Vanzella2023A&A...678A.173V,Nakajima2026Natur.653..363N}, AMORE6 at $z=5.73$ \citep[$3\sigma$ upper limit of $Z/Z_\odot<0.19\%$,][]{Morishita2025arXiv250710521M} and LAP2 at $z=4.19$ \citep[$2\sigma$ upper limit of $Z/Z_\odot<0.6\%$,][]{Vanzella2026A&A...705L..12V}, the little red dot A2744-QSO1 at $z=7.04$ with $4\times 10^{-3}Z_\odot$ \citep{Maiolino2026MNRAS.548f2109M}, the host galaxy of lensed SN~Eos at $z=5.13$ \citep[$Z/Z_\odot\lesssim1\%$,][]{Asada2026arXiv260714355A}, together with the serendipitously observed, unlensed MPG-CR3 at $z=3.19$ \citep[$2\sigma$ upper limit of $Z/Z_\odot<0.7\%$,][]{Cai2025ApJ...993L..52C}.

These record holders are, by construction, as close to pristine as current data allow, yet requiring a metal-free nebula may miss the longer-lived phase in which the first stars remain visible while their supernovae have already enriched the surrounding gas.
The ratio $\Rthree$ falls towards the lowest oxygen abundances, although it also depends on ionization parameter and spectral hardness in the low-metallicity regime \citep{Inoue2011MNRAS.415.2920I,Nakajima2022ApJS..262....3N,Sanders2024ApJ...962...24S}.
The NEFERTITI chemical-evolution models predict a brief self-polluted interval in which the first supernovae have already seeded the nebula with oxygen, producing $R3\sim1$ characteristic of the first enrichment episodes, while \piii stars still contribute more than half of the stellar mass \citep{Koutsouridou2023MNRAS.525..190K,Rusta2025ApJ...989L..32R}.
Faint, oxygen-weak dwarf galaxies at $z\lesssim5$ therefore sample the last epoch at which a first-enrichment episode could still be caught in the act.
This is exactly the epoch when the competing predictions for the end of \piii star formation diverge most sharply.

Photometric selection offers a route to such systems: strong \ha with weak \oiii$+$\hb imprints a distinctive pattern on JWST/NIRCam colours, and selections built on it have produced extremely metal-poor galaxy (EMPG) candidates at $z\sim4$--5 \citep{Nishigaki2023ApJ...952...11N}, the modestly lensed GLIMPSE-16043 at $z=6.20$ \citep{Fujimoto2025ApJ...989...46F, Fujimoto2025arXiv251211790F}, and further medium-band candidates at $2.5<z<6.5$ \citep{Trussler2026MNRAS.550g1360T}.
LATED (Lyman-Alpha Tomography of Extremely metal-poor Domains) is a systematic and generalized photometric selection framework for primordial galaxies and EMPGs from \lya emitters parent samples (Li et al., in preparation).
Starting from \lya emitters in the A2744 field with the VLT/MUSE lensing cluster project \citep{Richard2021A&A...646A..83R}, Li et al. (in preparation) selected A2744-11772 (RA, Dec. = 3.595913$^\circ$, -30.386383$^\circ$) at $z=4.80$ as an EMPG candidate.
Here we present serendipitous JWST integral field spectroscopy (IFS) observations to confirm its metal-poor nature.
As the first LATED candidate confirmed by spectroscopy obtained after its selection, we name it \target.
We adopt a flat $\Lambda$CDM cosmology with $H_0=70\kms\,\mathrm{Mpc^{-1}}$ and $\Omega_{\rm m}=0.3$, AB magnitudes, vacuum wavelengths, and a solar oxygen abundance $\logoh=8.69$ \citep{Asplund2021A&A...653A.141A}.

\section{Data and analysis}
\label{sec:data}

\begin{figure*}
\centering
\includegraphics[width=\textwidth]{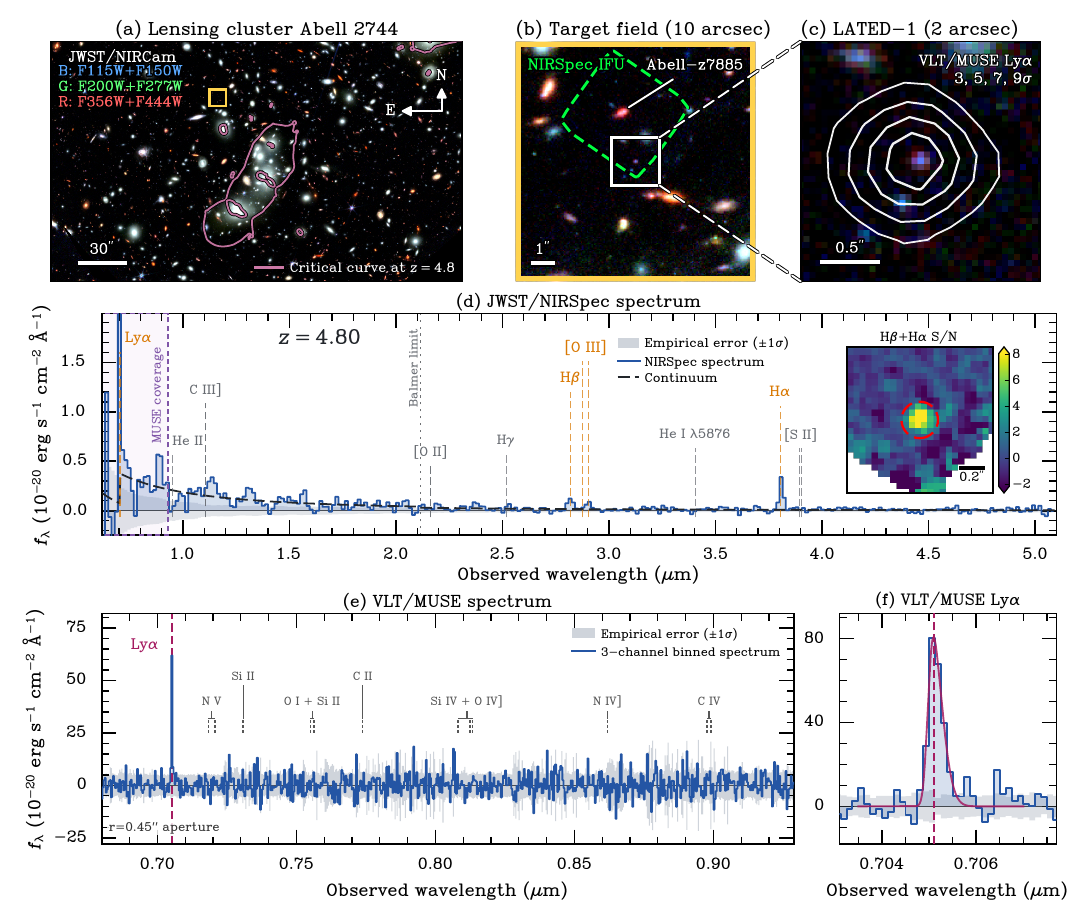}
\caption{JWST/NIRCam imaging, JWST/NIRSpec IFU and VLT/MUSE spectroscopy of \target.
(a) NIRCam colour composite of the Abell~2744 field, with the $z=4.80$ critical curve of the adopted lens model in pink and the target boxed in yellow.
(b) The 10-arcsec field, with the NIRSpec/IFU footprint outlined in green and the IFU central pointing target Abell-$z$7885 marked.
(c) The 2-arcsec field, with MUSE \lya narrow-band contours at 3, 5, 7, and 9$\sigma$.
(d) The NIRSpec/IFU spectrum with its empirical $1\sigma$ band and the fitted two-level continuum; the MUSE spectral coverage is shaded in purple, the lines are labelled, and the inset shows the \hb$+$\ha signal-to-noise map with the 0.15-arcsec extraction aperture.
(e) The VLT/MUSE spectrum, in which \lya is the only detection; grey ticks mark the other searched lines.
(f) The \lya profile at native resolution, with the skew-normal fit in magenta and its peak wavelength dashed.
}
\label{fig:context}
\end{figure*}

\subsection{JWST/NIRCam imaging, photometry, and lensing}
\label{sec:NIRCam}

We use the A2744 NIRCam mosaics presented in \citet{Fu2025ApJ...987..186F, Li2026arXiv260411892L}, which combine ERS-1324 (GLASS; PI: T. Treu; \citealt{Treu2022ApJ...935..110T}), GO-2561 (UNCOVER; PI: I. Labbé;  \citealt{Bezanson2024ApJ...974...92B}), DDT-2756 (PI: W. Chen), GO-2883 (MAGNIF; PI: F. Sun), GO-3516 (ALT; PI: J. Matthee; \citealt{Naidu2024arXiv241001874N}) and GO-4111 (MegaScience; PI: K. Suess; \citealt{Suess2024ApJ...976..101S}).
Together, these cover all eight NIRCam wide bands from F070W to F444W and twelve medium bands at the position of \target, with astrometry aligned to Gaia DR3 \citep{GaiaCollaboration2023A&A...674A...1G}.
Because \target is compact and faint, we measure fluxes within the resampled-PSF 50 per cent encircled-energy radius (Table~3 of the \href{https://jwst-docs.stsci.edu/jwst-near-infrared-camera/nircam-performance/nircam-point-spread-functions}{JWST NIRCam PSF documentation} and Table~\ref{tab:phot}) and apply the corresponding factor-of-two aperture correction.
The results are shown in Table~\ref{tab:phot}.

We adopt the UNCOVER strong-lens model v2.0 \citep{Furtak2023MNRAS.523.4568F, Price2025ApJ...982...51P}, which gives $\mu=2.80^{+0.03}_{-0.07}$ (statistical) and predicts no multiple images, placing the source well away from the critical curves (Figure~\ref{fig:context}a).
Independent Abell~2744 models scatter more than the statistical error \citep{Priewe2017MNRAS.465.1030P,Mahler2018MNRAS.473..663M,Bergamini2023A&A...670A..60B,Bergamini2023ApJ...952...84B}: evaluated at this position the public Frontier Fields models span $\mu=1.8$--$2.8$, with the free-form model of \citet{Vega-Ferrero2019MNRAS.486.5414V} reaching 5.5, and the MUSE lensing catalogue lists $\mu=1.99\pm0.04$ \citep{Richard2021A&A...646A..83R}.
However, these lens models are not the latest, as they are based on pre-JWST data or on very early JWST observations.
We therefore quote intrinsic quantities at the fiducial $\mu=2.80$ from the JWST UNCOVER model, but note a lower magnification would make \target both more intrinsically luminous and more massive.
Spectral energy distribution (SED) modelling of the magnification-corrected photometry with \textsc{Prospector} \citep{Johnson2021ApJS..254...22J}, with the redshift confined to within 0.03 of the spectroscopic value, gives stellar mass $\log(M_\star/M_\odot)=6.30^{+0.19}_{-0.16}$ and UV magnitude $M_{\rm UV}=-16.15\pm0.06$ at rest-frame 1500~\AA, where the intervals are 16th--84th percentile ranges and exclude SED-model and inter-model lensing systematics.
The same fit gives $t_{50}\simeq7.7$~Myr ($t_{90}\simeq1.3$~Myr), the lookback time by which 50 (90) per cent of the stellar mass had formed, and a star formation rate averaged over the last 10~Myr of $0.13^{+0.02}_{-0.01}\,M_\odot\,\mathrm{yr^{-1}}$.
Appendix~\ref{app:sed} gives the SED-modelling setup and results.

\subsection{Photometric selection with LATED methodology}

\target is selected by the LATED framework (Li et al., in preparation) to be an extremely metal-poor or metal-free candidate based on JWST NIRCam photometry.
LATED selection starts from \lya emitters with redshifts, because the redshift fixes which NIRCam bands contain the rest-frame optical lines.
\target is identified by MUSE observation to be a \lya emitter at $z=4.80$ (Section~\ref{sec:MUSE}), lying in the LATED W$z$4.4 window ($z=3.97$--4.86), in which \oiii$+$\hb falls in F277W, \ha in F356W, and the continuum redward of \ha in F444W.
LATED uses two criteria in the colour plane of $x\equiv m_\mathrm{F277W}-m_\mathrm{F356W}$ and $y\equiv m_\mathrm{F356W}-m_\mathrm{F444W}$ to perform the selection: $y\leq y_\mathrm{max}(z)$ selects a strong \ha excess over the continuum, tracing a high \ha equivalent width, and $x\geq x_\mathrm{min}(z)$ selects \ha dominance over \oiii$+$\hb, tracing a low oxygen abundance.
Both thresholds are the per-redshift envelope of the Pop~III templates of Yggdrasil \citep{Zackrisson2011ApJ...740...13Z} and the \textsc{cloudy} grids of \citet{Nakajima2022MNRAS.513.5134N}.
At $z=4.8$, they are $x_\mathrm{min}=0.147$ and $y_\mathrm{max}=-0.852$.
Both colours are flux ratios, so they do not depend on the lensing magnification.
Based on our photometry (Table~\ref{tab:phot}), \target has $x=0.42\pm0.19$ and $y=-1.29\pm0.47$, satisfying both criteria, with \ha-band F356W detected at $\mathrm{S/N}=10.0$.
These are consistent with measurements reported in Li et al. (in preparation) within $0.4\sigma$, which uses a different running-median background subtraction method designed for the crowded cluster wide field.

\subsection{JWST/NIRSpec IFU spectroscopy}

The source falls serendipitously in JWST GO-2957 observation~005 (PIs: H.~{\"U}bler \& R.~Maiolino, presented in \citealt{Koller2026MNRAS.551g1206K}), which comprises twelve NIRSpec/IFU PRISM cycling dithers obtained on 2024 August 1 for a total integration time of 17506.7~s.
The central target of the IFU observation is Abell-$z$7885, a main-sequence galaxy at $z=7.885$ \citep{Heintz2023NatAs...7.1517H}.
At $\sim2$~arcsec from the nominal pointing centre, \target lies near the dithered edge, so only seven dithers cover it at \hb$+$\oiii and six at \ha, giving effective exposure times of 10.2 and 8.8~ks rather than the full 17.5~ks.
We reduced the \texttt{rate} products, publicly available from the MAST archive, with the JWST Calibration Pipeline v2.0.1 and CRDS context \texttt{jwst\_1535.pmap} \citep{Bushouse2026zndo..20058613B,Greenfield2016A&C....16...41G}.
Pixels affected by failed-open shutters and exposure-specific artifacts were flagged, and the outermost illuminated border of every slice was removed before resampling at 0.05~arcsec per spaxel; a smooth spatial background was then fitted and subtracted in every wavelength plane with detected sources masked.
Finally, we registered the cube astrometry with a translation matching the \hb$+$\ha centroid of the target (inset of Fig.~\ref{fig:context}d) to the NIRCam coordinate (Fig.~\ref{fig:context}c).

\begin{figure}
\centering
\includegraphics[width=\columnwidth]{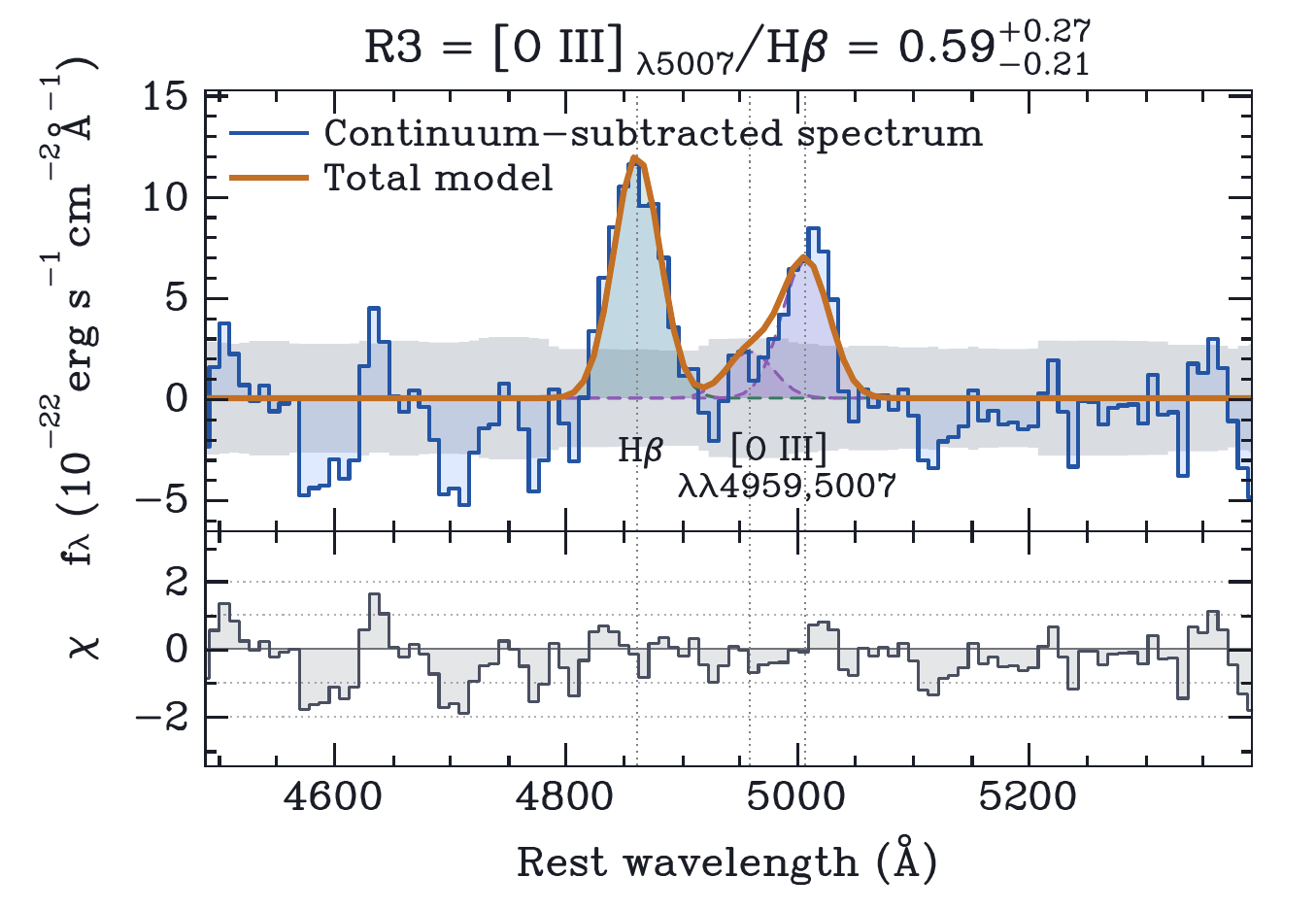}
\caption{The \hb$+$\oiii complex in the NIRSpec/IFU spectrum of \target.
The unbinned 0.15-arcsec-aperture continuum-subtracted spectrum and its $1\sigma$ band are plotted against rest-frame wavelength, together with the individual \hb, \oiii\,$\lambda4959$, and \oiii\,$\lambda5007$ Gaussians.}
\label{fig:spectrum}
\end{figure}

From the reduced cube, we extract a 0.15-arcsec radius source-centred aperture, subtract a local 0.30--0.55-arcsec annulus, and estimate the correlated-noise floor from $\sim100$ blank apertures, which is used to scale the \textsc{ERR} extension to get the final error, shown in Fig.~\ref{fig:context}d and Fig.~\ref{fig:spectrum}.
We then fit two independent, non-negative continuum levels (black dashed line in Fig.~\ref{fig:context}d), each flat in $f_\nu$ and split at the rest-frame Balmer limit, over the full PRISM range, attenuating the ultraviolet branch blueward of \lya by the IGM transmission of \citet{Inoue2014MNRAS.442.1805I}.
On the continuum-subtracted spectrum, we fit \hb and the \oiii doublet as three Gaussians sharing a common redshift and velocity width, with the doublet ratio fixed to $F_{\lambda5007}/F_{\lambda4959}=2.98$ \citep{Storey2000MNRAS.312..813S}, and \lya and \ha each as a single Gaussian.
We measure $F_\mathrm{H\alpha}=(80.8\pm4.6)\times10^{-20}$, $F_\mathrm{[O\,III]\lambda5007}=(21.1\pm7.4)\times10^{-20}$, $F_\mathrm{H\beta}=(35.8\pm6.4)\times10^{-20}$, and $F_\mathrm{Ly\alpha}=(418\pm60)\times10^{-20}\,\mathrm{erg\,s^{-1}\,cm^{-2}}$ without magnification correction, i.e. detections at 17.7, 2.9 (for \oiii doublet), 5.6, and 7.0$\sigma$.
The corresponding matched-aperture rest-frame equivalent widths are $\mathrm{EW}_0(\mathrm{Ly}\alpha)=192$~\AA, $\mathrm{EW}_0(\mathrm{H}\beta)=362$~\AA, and $\mathrm{EW}_0(\mathrm{H}\alpha)=1490$~\AA, with matched-blank 16th--84th percentile ranges of $152$--$241$, $256$--$605$, and $1122$--$2485$~\AA.
The aperture correction cancels between line and continuum, so it does not affect these values.
The \textsc{stpsf} model PSF encircled-energy fraction of the 0.15-arcsec aperture is 0.76 at \lya, 0.72 across \hb$+$\oiii, and 0.67 at \ha, so the total fluxes are $F_\mathrm{Ly\alpha}=(550\pm79)$, $F_\mathrm{H\beta}=(49.7\pm8.8)$, $F_\mathrm{[O\,III]\lambda5007}=(29.3\pm10.3)$, and $F_\mathrm{H\alpha}=(120.5\pm6.8)\times10^{-20}\,\mathrm{erg\,s^{-1}\,cm^{-2}}$.
Only \ha has sufficient S/N to test the correction independently: its 0.15--0.25-arcsec curve of growth reproduces the same total and sets a systematic term consistent with a point source, giving $F_\mathrm{H\alpha}=(120.5\pm6.8_{\rm stat}\pm4.0_{\rm sys})\times10^{-20}\,\mathrm{erg\,s^{-1}\,cm^{-2}}$, whereas the other lines are too faint for the curve of growth test.
The non-resonant lines give \zprism, where the interval is the \hb$+$\oiii-to-\ha scatter and is dominated by the known PRISM wavelength calibration rather than by photon noise \citep[e.g.,][]{deGraaff2025A&A...697A.189D}.
We also constrain \heii\,$\lambda1640$ and \heii\,$\lambda4686$ through source-position injection and recovery tests, obtaining \HeIIUVResult and \HeIIOptResult without aperture correction.

In Appendix~\ref{app:halpha_mass}, we show that the \ha luminosity independently requires $\log(M_\star/M_\odot)=4.8$--6.0 for a Pop~III population with various initial mass functions (IMFs) and $\log(M_\star/M_\odot)=6.1$--6.4 for a young Pop~II population with canonical IMFs \citep{Chabrier2003PASP..115..763C,Kroupa2001MNRAS.322..231K,Salpeter1955ApJ...121..161S}.

The \hb and \oiii fluxes yield $\Rthree={}$\RthreeResult (Fig.~\ref{fig:spectrum}).
We performed a series of quality assessments on this measurement.
Compact 0.10--0.20-arcsec apertures give $R3=0.56$--0.71 and jackknife resampling (leave one exposure out) stacks span 0.50--0.74.
Injected \oiii lines of the measured fluxes are recovered above $2\sigma$ in 86 per cent of trials with a 1.1 per cent false-positive rate.
The fixed 0.15-arcsec-aperture Balmer decrement is $\mathrm{H}\alpha/\mathrm{H}\beta=2.26\pm0.42$, while aperture-corrected total fluxes give $2.43\pm0.45$, consistent with typical Case-B condition within $1\sigma$.
Because \hb enters both the Balmer decrement and $R3$, a noise-high \hb would both lower $R3$ and decrease the decrement together; replacing the measured \hb by the value implied by the 17.7$\sigma$ \ha under Case B (\ha/\hb=2.86) is therefore a conservative estimate for the result, and it still gives $R3=0.69$.

\subsection{VLT/MUSE spectroscopy}
\label{sec:MUSE}

The VLT/MUSE observations are taken by the MUSE Lensing Cluster project (PI: Richard, \citealt{Richard2021A&A...646A..83R}).
We registered the cube to the Gaia-DR3-aligned JWST NIRCam F090W mosaic with a similarity transform fitted to 108 continuum sources common to F090W and an 8000--9200~\AA\ collapse of the cube, leaving a 43 mas residual RMS.
We then extracted a 0.45-arcsec-radius aperture spectrum (Fig.~\ref{fig:context}e) at the registered JWST position, estimating the noise empirically from blank apertures of matched radius placed 2.5--7.0~arcsec away after masking catalogue sources.
The MUSE wavelength axis is converted from air to vacuum \citep{Birch1993Metro..30..155B, Birch1994Metro..31..315B, Morton2000ApJS..130..403M}.

\lya is the only line significantly detected ($>3\sigma$) in the extracted spectrum, after a systematic line detection search.
Because \lya is resonantly scattered and red-asymmetric, we fit its profile with a skew-normal function over 7035--7070~\AA\ rather than a Gaussian, propagating the empirical errors through Monte Carlo realisations.
This gives a peak wavelength of $7050.95\pm0.26$~\AA\ ($\sigma=2.48$~\AA, skewness parameter $\gamma=2.95$, shown in Fig.~\ref{fig:context}f), i.e. \zlya.
For the line flux we adopt the catalogue value $(5.35\pm0.61)\times10^{-18}\,\mathrm{erg\,s^{-1}\,cm^{-2}}$ \citep{Richard2021A&A...646A..83R}, which comes from a segmentation-weighted optimal extraction rather than a fixed aperture and is therefore not directly comparable to our own 0.45-arcsec aperture, which contains half as much flux.
The peak bounds the systemic redshift from above and agrees with \zprism, so we adopt $z=4.80$ throughout.
The \lya contours in Fig.~\ref{fig:context}c come from a continuum-subtracted narrow band over 7046.6--7054.1~\AA.
The point-source-corrected NIRSpec \lya flux agrees with the adopted MUSE value (ratio $1.03\pm0.19$), so the two instruments are consistent.

We note that the \citet{Richard2021A&A...646A..83R} catalogue reports \civ$\lambda1548$ at S/N of 3.6, but that fit simultaneously reports its $\lambda1551$ partner at $(8.6\pm30.6)\times10^{-20}$, an unphysical $10{:}1$ ratio, whereas the intrinsic ratio set by the oscillator strengths is $2{:}1$ and resonant scattering and absorption preferentially suppresses $\lambda1548$ \citep{Berg2019ApJ...878L...3B, Topping2024MNRAS.529.3301T}.
Re-extracting the doublet over nine aperture radii from 0.20 to 1.20~arcsec recovers \lya but never reaches $2\sigma$ in \civ, and freeing both amplitudes returns a negative $\lambda1551$ component at every compact aperture.
We therefore treat \civ as undetected.
Combining the 0.45-arcsec MUSE measurement with the NIRSpec PRISM spectrum by weighting the two by inverse variance, each corrected to total flux, gives $F_\mathrm{C\,{IV}\lambda\lambda1548,1551}<1.2\times10^{-18}\,\mathrm{erg\,s^{-1}\,cm^{-2}}$ and $F_\mathrm{C\,{IV}\lambda\lambda1548,1551}/F_\mathrm{Ly\alpha}<0.23$ at $2\sigma$.

\begin{figure*}
\centering
\includegraphics[width=\textwidth]{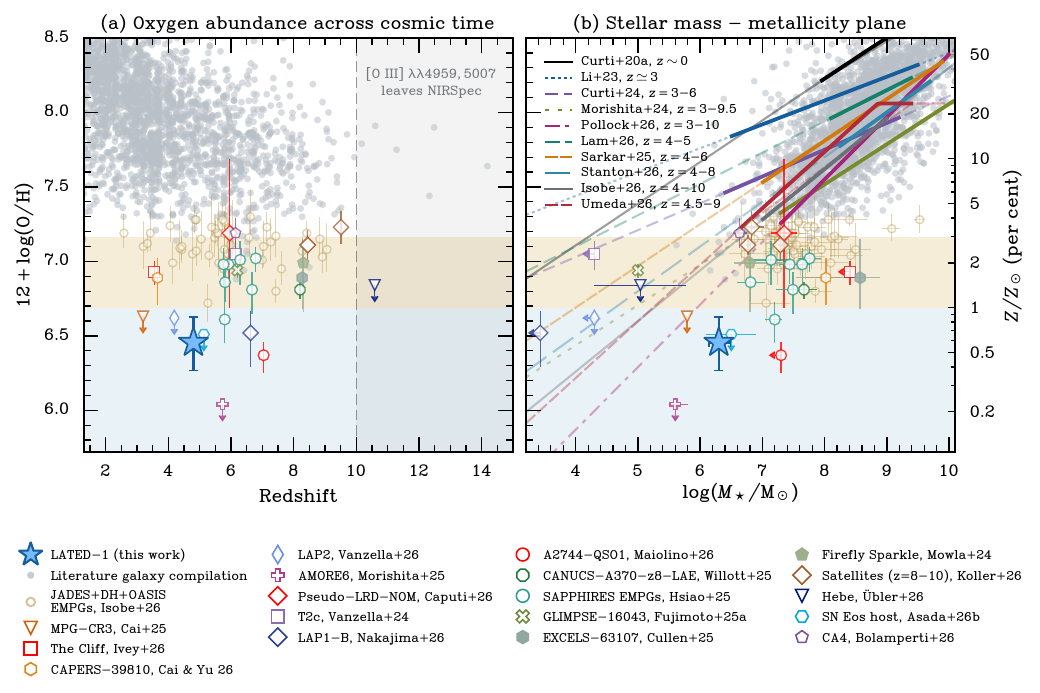}
\caption{Oxygen abundance of \target against redshift (a) and stellar mass (b), on a shared oxygen abundance axis with the corresponding solar fraction on the right.
Other spectroscopically confirmed extremely metal-poor systems are placed, like \target, on the single low-metallicity calibration of \citet{Isobe2026arXiv260611345I} so that the frontier objects are directly comparable; the resulting abundances therefore differ from those in the discovery papers.
Colour and marker identify each object or survey, two green filled symbols denote direct-$T_{\rm e}$ abundances, and arrows are $2\sigma$ upper limits.
The three little red dots are the exception and we keep their published values, because the $[\mathrm{O\,III}]$ strong-line calibrations may not apply at their inferred gas densities; the pseudo-LRD bar is a photoionization-model range rather than an uncertainty \citep{Caputi2026ApJ..1007..203C}.
Pale grey points are $\sim$2400 literature galaxies at $z=1.0$--14.2 (Appendix~\ref{app:r3}).
Curves in panel (b) are published mass--metallicity relations over the labelled redshift ranges \citep{Li2023ApJ...955L..18L,Curti2024A&A...684A..75C,Morishita2024ApJ...971...43M,Pollock2026A&A...708A.203P,Lam2026arXiv260530513L,Sarkar2025ApJ...978..136S,Stanton2026MNRAS.547.stag449S,Isobe2026arXiv260611345I,Umeda2026arXiv260715515U}, with the local relation of \citet{Curti2020MNRAS.491..944C} in black, solid over the stellar-mass range each fit covers and dashed where extrapolated.
The yellow band marks 1--3 per cent solar and lightblue band marks $<1$ per cent solar.}
\label{fig:models}
\end{figure*}

\section{A dwarf galaxy caught in the first chemical enrichment}
\label{sec:interpretation}

Applying $\Rthree=0.59^{+0.27}_{-0.21}$ through the low-metallicity extension of the \citet{Isobe2026arXiv260611345I} strong line calibration gives $\logoh=6.45^{+0.17}_{-0.19}$, or $0.58^{+0.28}_{-0.20}$ per cent of the solar oxygen abundance.
The $R3$ calibration is formally double-valued, but the metal-rich end predicts $R2\equiv[\mathrm{O\,II}]\lambda\lambda3726,3729/\mathrm{H}\beta\simeq2.8$ and $R23\equiv([\mathrm{O\,II}]\lambda\lambda3726,3729+[\mathrm{O\,III}]\lambda\lambda4959,5007)/\mathrm{H}\beta\simeq3.4$ against our $2\sigma$ limits of $0.38$ and $1.17$.
The low-metallicity branch solution is independently supported by its intrinsically faint UV magnitude $M_{\rm UV}=-16.15$, low stellar mass ($\log(M_\star/M_\odot)=6.30^{+0.19}_{-0.16}$), and extremely high line EWs ($\mathrm{EW}_0(\mathrm{H}\beta)=362\, \AA$ and $\mathrm{EW}_0(\mathrm{H}\alpha)=1490$~\AA).
The same $R23$ limit removes the intermediate-metallicity, low-ionization solution, because $R23$ stays above unity throughout $7.0\lesssim\logoh\lesssim8.5$ at any reasonable ionization parameter.

The unignorable uncertainty comes from the strong-line calibration.
\citet{Isobe2026arXiv260611345I} anchor their stacks empirically over $7.0\leq\logoh\leq8.7$, and below $7.0$ fix the slope of every index to \textsc{Cloudy} models with BPASS ionizing spectra, warning that abundances outside the calibrated range carry significant systematics from both the calibration and the extrapolation itself.
Our value lies $0.55$~dex below that boundary, and $0.27$~dex below the most metal-poor of their $50$ candidates.
Alternative calibrations quantify this systematic spread (Appendix~\ref{app:r3}): extrapolating \citet{Nakajima2022ApJS..262....3N} gives $\logoh=6.63$, and the relation of \citet{Sanders2024ApJ...962...24S} gives $\logoh=6.38^{+0.13}_{-0.14}$.
That $0.24$~dex range is comparable to the statistical interval and consistent with the ${>}0.3$~dex spread reported elsewhere at these abundances \citep{Cai2025ApJ...993L..52C}.
Photoionization models tie the inferred abundance to the ionization parameter: it stays below one per cent solar for $\log U\gtrsim-2.3$ (Appendix~\ref{app:r3}).
One property of the source works in its favour, since \citet{Nakajima2022ApJS..262....3N} find $R3$ alone accurate to $0.10$~dex for local EMPGs with $\mathrm{EW}_0(\mathrm{H}\beta)>200$~\AA, and at $362$~\AA\ \target sits well inside that regime.
We therefore quote $\logoh=6.45^{+0.17}_{-0.19}$ as a fiducial model-anchored strong-line inference in this work.

Figure~\ref{fig:models} places \target in context with other spectroscopically confirmed extremely metal-poor systems reduced to the same calibration as we use and uniformly adopting 2$\sigma$ upper limits when \oiii is not detected.
\target is among the most metal-poor galaxies known at any redshift, with an oxygen abundance below $1\%$ of the solar value.
LAP1-B and the little red dot A2744-QSO1 sit in a similar regime, and AMORE6, the pseudo-LRD-NOM, the host of SN~Eos, LAP2, and MPG-CR3 remain upper limits near or below our value \citep{Nakajima2026Natur.653..363N,Morishita2025arXiv250710521M,Maiolino2026MNRAS.548f2109M,Caputi2026ApJ..1007..203C,Asada2026arXiv260714355A,Vanzella2026A&A...705L..12V,Cai2025ApJ...993L..52C}.
On the mass-metallicity plane, \target separates from the population, and the separation is already present in the measured $R3$ (Appendix~\ref{app:r3}).
Extrapolating the $z=3$--6 relation of \citet{Curti2024A&A...684A..75C} to $\log(M_\star/M_\odot)=6.30$ predicts $\logoh=7.44$, so \target falls $1.0$~dex below it.
More tellingly, this is not an extrapolation for every sample.
\citet{Asada2026arXiv260120045A} measure individual galaxies down to $\log(M_\star/M_\odot)=5.6$ with deep NIRSpec observations, and their least massive object is $0.7$~dex more metal-rich than \target at $0.7$~dex lower mass.
\target therefore sits below the mass--metallicity relation in the regime where that relation is directly constrained.
The size of the deficit depends on which relation is used, spanning $0.4$--$1.4$~dex across the ten plotted with a median of $0.81$~dex, but it is robust against all of them.
\citet{Asada2026arXiv260120045A} measure an intrinsic scatter of $0.23$~dex over $10^{5.6}$--$10^{7.0}\,M_\odot$, spanning \target's mass, so even the smallest offset is $\simeq1.7\sigma$ and the median $\simeq3.5\sigma$.
\target has therefore not acquired the metals its stellar mass implies, as expected of a galaxy caught before self-enrichment has run its course.
Such a deficit arises on the delayed-enrichment (`undershoot') path of the A-SLOTH model \citep{Liu2025arXiv250606139L} shown by \citet{Asada2026arXiv260120045A}, in which a halo stays nearly pristine until a major burst of star formation, although at that path merges with the mass--metallicity relation above $M_\star\approx10^{6}\,M_\odot$ $z\approx6$, so \target would sit at its massive end.
The same path has been proposed for the host of SN~Eos, with $M_\star\approx10^{6.5}\,M_\odot$ and $\logoh<6.67$ ($2\sigma$) at $z=5.13$ \citep{Asada2026arXiv260714355A}.

The \piii timescales point the same way.
The most massive \piii stars that eject metals, of 140--260~$M_\odot$, explode as pair-instability supernovae after only 2.1--2.4~Myr \citep{Heger2002ApJ...567..532H,Schaerer2002A&A...382...28S}, earlier than any core-collapse supernova of a normal-metallicity population \citep[$\simeq3$--4~Myr;][]{Eldridge2017PASA...34...58E}, so a \piii population can enrich its own gas within a few million years.
\piii models reproduce the observed \ha and UV luminosities together at about 2--3~Myr if roughly half of the ionizing photons escape ($f_\mathrm{cov}=0.5$), or at 6--10~Myr if $f_\mathrm{cov}=1$ (Appendix~\ref{app:halpha_mass}), around or after the time its most massive members explode.
The NEFERTITI models \citep{Koutsouridou2023MNRAS.525..190K} describe this configuration as a self-polluted phase in which the first supernovae have enriched the gas while \piii stars still dominate the stellar mass, and predict $R3\approx1$ for it in galaxies at $z=6$--17 \citep{Rusta2025ApJ...989L..32R}.
The phase is brief, unresolved at their $1$~Myr sampling, so few galaxies should ever be caught in it, and \target has an $R3$ and abundance consistent with it.
The \ha luminosity adds a condition: a \piii population younger than 2~Myr with covering fraction of $f_\mathrm{cov}=1$ needs $10^{4.8}$--$10^{6.0}\,M_\odot$ to power all of the \ha, depending on the IMF (Appendix~\ref{app:halpha_mass}).
This is 3--50 per cent of the \textsc{Prospector} mass, which assumes a Pop~II population with a canonical \citet{Chabrier2003PASP..115..763C} IMF.
The same \ha can therefore come from a \piii-dominated system of $10^{4.8}$--$10^{6.8}\,M_\odot$, depending on its IMF, age and photon escape, or from $\simeq10^{6.3}\,M_\odot$ of Pop~II stars, and its luminosity alone cannot distinguish the two.
Rotation and binary interactions, which the non-rotating single-star \piii models used here omit, can prolong the lives of \piii stars and so keep their ionizing output high for longer \citep{Yoon2012A&A...542A.113Y,Murphy2021MNRAS.501.2745M,Sibony2022A&A...666A.199S,Lecroq2025A&A...695A..17L,Wasserman2026MNRAS.547ag386W}.
In addition, rapid rotation of \piii stars can trigger chemically homogeneous evolution that can further boost the ionizing flux (and \ha emission per unit stellar mass formed) of \piii stars by a factor of 2--6 \citep{Liu2025MNRAS.541.3113L}.
All of these effects lower the stellar mass inferred from H$\alpha$ in the \piii scenario.

The methodological result is independent of interpretation.
\target was selected as an extremely metal-poor candidate by LATED (Li et al., in preparation) from NIRCam photometry, which predicted $R3<1.54$ at $2\sigma$ before any spectroscopic oxygen constraint existed, and this JWST IFU spectrum returned an abundance below one per cent solar.
We would like to highlight that the comparison cases show how seldom this has happened.
GLIMPSE-16043 was photometrically predicted to have $R3<0.44$ ($1\sigma$) and $Z/Z_\odot<0.5\%$, but follow-up NIRSpec medium-resolution spectroscopy returned $R3=1.78\pm0.18$ instead \citep{Fujimoto2025arXiv251211790F,Fujimoto2025ApJ...989...46F}.
Of the $22$ medium-band candidates of \citet{Trussler2026MNRAS.550g1360T}, whose photometric signature closely resembles ours, the one with NIRSpec coverage is confirmed metal-poor but at $\logoh=7.19$ with a strong Balmer decrement, above the metallicity threshold on which it was selected \citep{Isobe2026arXiv260611345I}.
Photometric pre-selection is well established locally \citep{Kojima2020ApJ...898..142K,Nishigaki2023ApJ...952...11N}; what has been missing is a high-redshift candidate that survived spectroscopy at the depth at which it was selected.

This matters because the present frontier is built from objects that were not searched for systematically.
LAP1-B reaches $\logoh=6.31$ at magnification of $\mu\simeq98$ with $M_\star<3300\,M_\odot$, and AMORE6, behind the same cluster as \target, is more pristine still \citep{Nakajima2026Natur.653..363N,Morishita2025arXiv250710521M}; both were found by examining known caustic arcs.
\target has an $R3$ indistinguishable from LAP1-B's $0.69\pm0.28$ but at moderate $\mu=2.80$, so its intrinsic properties do not depend on an extreme lens model, and among galaxies rather than star-cluster-scale complexes, its abundance is among the lowest reported.
A selection that works at ordinary magnification can be run over survey volumes, which turns single objects into a population and the apparent $1$--2 per cent solar floor into a testable statement rather than a description of which galaxies get followed up \citep{Hsiao2025arXiv250503873H,Isobe2026arXiv260611345I}.
The redshift sharpens the test, because simulations that end primordial star formation with reionization and semi-analytic models that let it persist diverge most sharply near $z\simeq5$ \citep{Zier2025MNRAS.544..410Z, Liu2020MNRAS.497.2839L}.
Although \target alone could not fully settle that question, it shows that a galaxy in the earliest state of chemical enrichment can be found by design rather than by luck, and that at least a few such systems survive to $z<5$.
Each such galaxy moves the first stars from an inference about when they ended towards a population that can be found and counted.

\section{Summary}
\label{sec:conclusions}

We have presented serendipitous JWST/NIRSpec IFU spectroscopy of \target, an intrinsically faint \lya emitter at $z=4.80$ behind Abell~2744 that LATED had selected from NIRCam imaging as an extremely metal-poor or metal-free candidate.
The spectrum shows \lya at a flux consistent with MUSE, spatially coincident \hb and \ha, and only a marginal \oiii doublet, giving $R3={}$\RthreeResult.
The ratio is stable across compact apertures and jackknife resampling stacks, and a systematic line search finds no further secure ultraviolet line.
With the metal-rich branch of R3 diagnostic excluded by the \oii and $R23$ limits, the adopted calibration gives $\logoh=6.45^{+0.17}_{-0.19}$, or $0.58^{+0.28}_{-0.20}$ per cent of the solar oxygen abundance.
That places \target among the most metal-poor galaxies known and about $1$~dex below the mass--metallicity relation at its stellar mass at the same redshift.
Unlike the caustic-magnified complexes that reach comparable abundances, its intrinsic properties do not depend on an extreme lens model.

The abundance remains a model-anchored strong-line inference below the empirical calibration floor, and \heii is undetected.
The measurement establishes that such a galaxy can be picked out from imaging before any spectroscopic constraint exists: to our knowledge \target is the first photometrically pre-selected candidate whose follow-up spectroscopy has returned an abundance below one per cent solar.
Chemically primitive conditions therefore survive to $z<5$ in galaxies that ordinary surveys can reach, which turns the end of primordial star formation from a modelling question into a searchable one.
Deeper medium-resolution spectroscopy of \heii, \oiii$\lambda4363$, and the C/O-sensitive ultraviolet lines would settle both the abundance and the stellar population that powers the nebula.

Because the IFU pointing targeted an unrelated source, these data serve as a blind test of LATED.
Both LATED-selected galaxies in Abell~2744 with rest-frame optical spectroscopy, \target and the independently recovered AMORE6, have returned abundances below one per cent solar (Li et al., in preparation).
LATED thus offers an efficient selection route to extremely metal-poor galaxies, and to candidate \piii or self-polluted \piii systems.

\section*{Acknowledgements}
RM acknowledges support from the Science and Technology Facilities Council (STFC), the European Research Council (ERC) through Advanced Grant 695671 ``QUENCH'', and the UK Research and Innovation (UKRI) Frontier Research grant RISEandFALL.
RM also acknowledges support from a Royal Society Research Professorship grant. 
ZC acknowledges support from National Key R\&D
Program of China (grant No. 2023YFA1605600), National Natural Science Foundation of China (\#12525303), Tsinghua University Initiative Scientific Research Program, and New Cornerstone Science Foundation through the XPLORER PRIZE. 
H\"U acknowledges support by the Max Planck Society through the Lise Meitner Excellence Program. H\"U acknowledges funding by the European Union (ERC APEX, 101164796). Views and opinions expressed are however those of the authors only and do not necessarily reflect those of the European Union or the European Research Council Executive Agency. Neither the European Union nor the granting authority can be held responsible for them.
BL acknowledges financial support from the German Excellence Strategy via the Heidelberg Cluster of Excellence (EXC 2181 -- 390900948) STRUCTURES. 
This work is based in part on observations made with the NASA/ESA/CSA James Webb Space Telescope. The data were obtained from the Mikulski Archive for Space Telescopes at the Space Telescope Science Institute, which is operated by the Association of Universities for Research in Astronomy, Inc., under NASA contract NAS 5-03127 for JWST. These observations are associated with program \#1324, \#2561, \#2756, \#2883, \#2957, \#3516, \#3538 and \#4111.
The authors acknowledge the teams of JWST programs for developing their observing program with a zero-exclusive-access period.
Based on observations collected at the European Organisation for Astronomical Research in the Southern Hemisphere under ESO programme 094.A-0115.
The JWST data described here may be obtained from the MAST archive at
\url{https://dx.doi.org/10.17909/txr8-xm89}.
Generative AI (OpenAI ChatGPT and Anthropic Claude) assisted with draft language editing, coding, and literature searching. The authors reviewed and approved all scientific content, interpretations, references, and final wording and take full responsibility for the manuscript.

\section*{Data availability}

The data used in this work are publicly available from MAST and the MUSE data release presented by \citet{Richard2021A&A...646A..83R}.
The literature measurements plotted in Figs~\ref{fig:models} and \ref{fig:r3cal} are taken from the publications shown in Appendix~\ref{app:r3}.

\bibliographystyle{mnras}
\bibliography{main}

\vspace{4mm}
\printaffils

\appendix

\section{Prospector SED fitting}
\label{app:sed}

We fit the NIRCam photometry of \target (Table~\ref{tab:phot}) with \textsc{Prospector} \citep{Johnson2021ApJS..254...22J}.
The data are the EE50 aperture fluxes of Section~\ref{sec:data} in the eight wide and twelve medium bands from F070W to F480M, divided by the fiducial magnification $\mu=2.80$, so the posteriors exclude the lensing uncertainty.
Low-S/N bands enter as forced flux measurements rather than upper limits, and the flux uncertainties have a floor of 5 per cent of the flux.
The stellar populations are computed with FSPS \citep{Conroy2009ApJ...699..486C,Conroy2010ApJ...712..833C}, using the MIST isochrones \citep{Choi2016ApJ...823..102C,Dotter2016ApJS..222....8D}, the MILES spectral library \citep{Sanchez-Blazquez2006MNRAS.371..703S,Falcon-Barroso2011A&A...532A..95F}, and a \citet{Chabrier2003PASP..115..763C} IMF.

\begin{table}
\centering
\caption{JWST photometry of \target.
Fluxes are measured within the 50 per cent encircled-energy radius $r_\mathrm{EE50}$ of each filter, after subtracting a local background fitted in a 0.6--1.0~arcsec annulus, and corrected to total flux by a factor of two.
They are observed, image-plane flux densities; the fit uses them divided by $\mu=2.80$ with a 5 per cent error floor.}
\label{tab:phot}
\renewcommand{\arraystretch}{1.1}
\begin{tabular}{lccr}
\toprule
Filter & $r_\mathrm{EE50}$ [arcsec] & $f_\nu$ [nJy] & S/N \\
\midrule
F070W & 0.048 & $13.86\pm2.31$ & 6.0 \\
F090W & 0.047 & $9.11\pm1.11$ & 8.2 \\
F115W & 0.048 & $8.24\pm0.88$ & 9.3 \\
F140M & 0.051 & $8.21\pm1.78$ & 4.6 \\
F150W & 0.051 & $6.86\pm0.90$ & 7.6 \\
F162M & 0.052 & $5.19\pm1.73$ & 3.0 \\
F182M & 0.054 & $6.97\pm1.28$ & 5.4 \\
F200W & 0.055 & $5.41\pm0.76$ & 7.2 \\
F210M & 0.056 & $4.17\pm1.05$ & 4.0 \\
F250M & 0.091 & $1.52\pm3.44$ & 0.4 \\
F277W & 0.094 & $7.40\pm1.07$ & 6.9 \\
F300M & 0.097 & $10.04\pm2.51$ & 4.0 \\
F335M & 0.102 & $4.17\pm2.23$ & 1.9 \\
F356W & 0.105 & $10.87\pm1.09$ & 10.0 \\
F360M & 0.105 & $11.50\pm2.46$ & 4.7 \\
F410M & 0.110 & $5.75\pm2.30$ & 2.5 \\
F430M & 0.112 & $3.68\pm5.15$ & 0.7 \\
F444W & 0.113 & $3.33\pm1.42$ & 2.3 \\
F460M & 0.116 & $-3.14\pm6.19$ & $-$0.5 \\
F480M & 0.118 & $10.06\pm9.46$ & 1.1 \\
\bottomrule
\end{tabular}
\end{table}

We use the flexible non-parametric star formation history (SFH), the Continuity SFH \citep{Leja2019ApJ...876....3L}, with a constant star formation rate in each of ten bins of lookback time: 0--3 and 3--10~Myr, followed by eight bins logarithmically spaced up to the time elapsed since $z=20$ ($\approx1$~Gyr).
The logarithmic ratios of the star formation rates in adjacent bins follow the non-uniform star-forming main sequence prior of \citet{Duan2026arXiv260521599D}: Student's $t$ distributions with two degrees of freedom and a scale of 0.5, truncated at $\pm5$ and centred on the ratios obtained by integrating the redshift-dependent star-forming main sequence backward in time from the sampled redshift and mass.
We use flat priors on total mass formed and stellar metallicity over $4<\log(M_\mathrm{formed}/M_\odot)<12$ and $-2.63<\log(Z_\star/Z_\odot)<-0.5$, and a flat prior on redshift within $\pm0.03$ of the spectroscopic value.
Dust attenuation follows the two-component model of \citet{Charlot2000ApJ...539..718C}.
The diffuse optical depth $\hat{\tau}_\mathrm{dust,2}$ attenuates all stars with the curve of \citet{Kriek2013ApJ...775L..16K}, in which a free power-law modifier $n$ tilts the \citet{Calzetti2000ApJ...533..682C} curve by a factor $(\lambda/5500\,\AA)^{n}$ and sets the strength of the 2175~\AA\ bump.
Stars younger than 10~Myr are further attenuated by birth clouds with optical depth $\hat{\tau}_\mathrm{dust,1}$.
Nebular continuum and line emission, including \lya, come from the \textsc{cloudyfsps} \citep{Byler2024ascl.soft09008B}, \textsc{Cloudy}-based grids of \citet{Byler2017ApJ...840...44B}, with the gas-phase metallicity and ionization parameter left free and independent of the stellar metallicity, and the IGM attenuation of \citet{Madau1995ApJ...441...18M} is applied with a fixed normalization.
Table~\ref{tab:sed} lists the priors of all 17 free parameters.
The posterior is sampled with \textsc{nautilus} \citep{Lange2023MNRAS.525.3181L} using 2000 live points until an effective sample size of 5000 is reached.

\begin{table}
\centering
\caption{Priors and posteriors of the \textsc{Prospector} fit, with posterior medians and 16th--84th percentile ranges.
$\mathcal{U}$ denotes a flat prior and $\mathcal{N}$ a normal prior truncated to the range given; quantities below the line are derived from the posterior of free parameters.}
\label{tab:sed}
\renewcommand{\arraystretch}{1.2}
\begin{tabular}{lcc}
\toprule
Parameter & Prior & Posterior \\
\midrule
$z$ & $\mathcal{U}(4.770,4.830)$ & $4.805^{+0.013}_{-0.016}$ \\
$\log(M_\mathrm{formed}/M_\odot)$ & $\mathcal{U}(4,12)$ & $6.33^{+0.22}_{-0.17}$ \\
$\log(Z_\star/Z_\odot)$ & $\mathcal{U}(-2.63,-0.5)$ & $-2.06^{+0.82}_{-0.44}$ \\
$\log(\mathrm{SFR}_i/\mathrm{SFR}_{i+1})$, 9 ratios & SFMS prior & Fig.~\ref{fig:sed}c \\
$\hat{\tau}_\mathrm{dust,2}$ & $\mathcal{N}(0,0.5)$, $[0,4]$ & $0.04^{+0.03}_{-0.02}$ \\
$n$ & $\mathcal{U}(-1,0.4)$ & $-0.01^{+0.31}_{-0.45}$ \\
$\hat{\tau}_\mathrm{dust,1}/\hat{\tau}_\mathrm{dust,2}$ & $\mathcal{N}(1,0.3)$, $[0,2]$ & $0.93\pm0.29$ \\
$\log(Z_\mathrm{gas}/Z_\odot)$ & $\mathcal{U}(-2,0.5)$ & $-1.47^{+0.29}_{-0.33}$ \\
$\log U$ & $\mathcal{U}(-4,-1)$ & $-1.40^{+0.29}_{-0.42}$ \\
\midrule
$\log(M_\star/M_\odot)$ & & $6.30^{+0.19}_{-0.16}$ \\
$\mathrm{SFR}_{10}$ [$M_\odot\,\mathrm{yr^{-1}}$] & & $0.13^{+0.02}_{-0.01}$ \\
$\mathrm{SFR}_{100}$ [$M_\odot\,\mathrm{yr^{-1}}$] & & $0.020^{+0.012}_{-0.006}$ \\
$t_{50}$ [Myr] & & $7.7^{+10.5}_{-5.0}$ \\
$t_{90}$ [Myr] & & $1.3^{+2.1}_{-0.7}$ \\
$M_\mathrm{UV}$ & & $-16.15\pm0.06$ \\
$M_\mathrm{UV}$, dust-free & & $-16.40\pm0.10$ \\
\bottomrule
\end{tabular}
\end{table}

\begin{figure*}
\centering
\includegraphics[width=\textwidth]{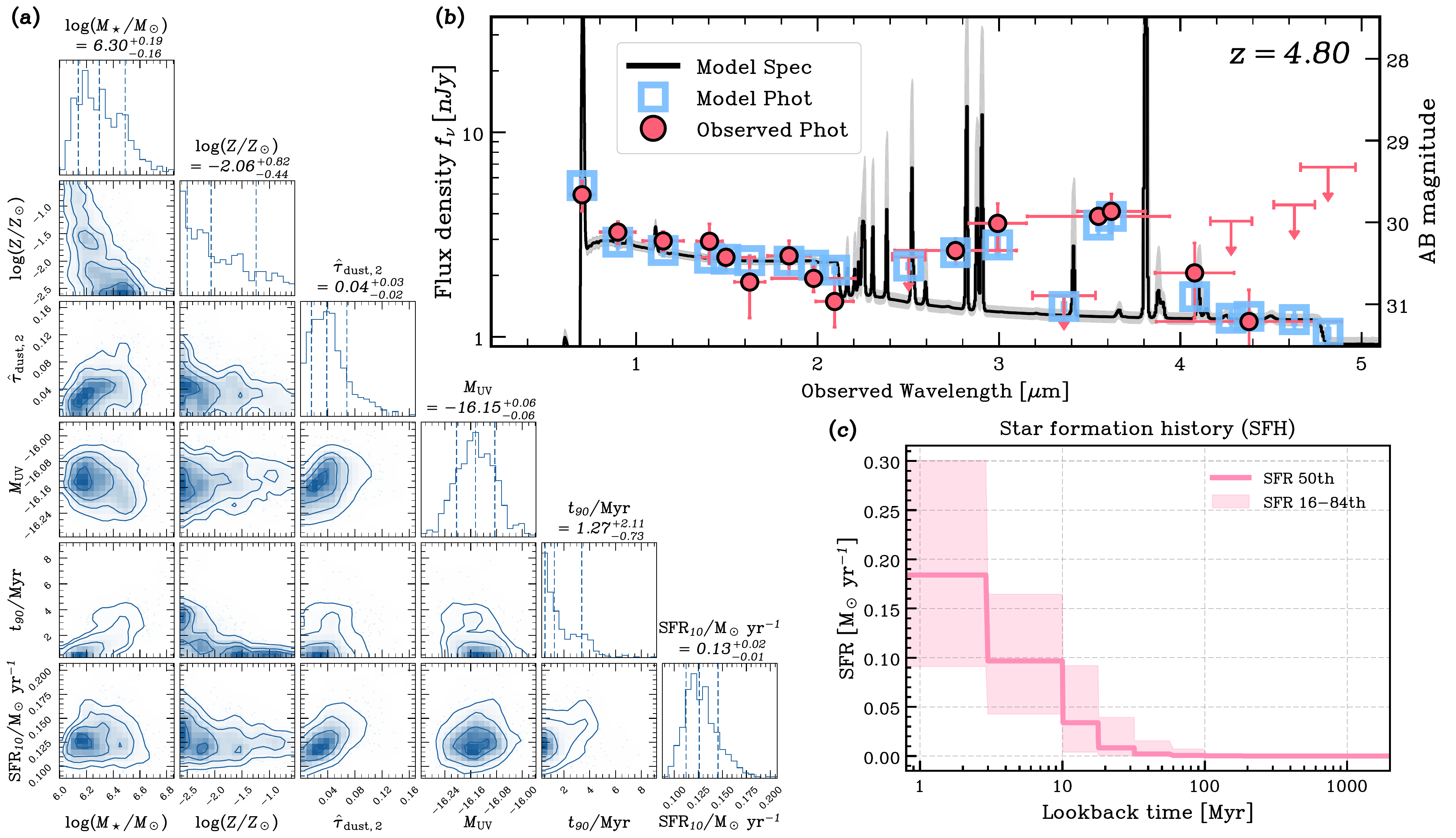}
\caption{\textsc{Prospector} SED fit of \target.
(a) Posterior distributions of the stellar mass, stellar metallicity, diffuse dust optical depth, $M_{\rm UV}$, $t_{90}$, and star formation rate averaged over the last 10~Myr, with dashed lines at the 16th, 50th, and 84th percentiles.
(b) Magnification-corrected NIRCam photometry (red circles, with arrows marking $2\sigma$ upper limits for bands below $2\sigma$), median model photometry (blue squares), and the median model spectrum (black) with its 16th--84th percentile range (grey).
(c) Median star formation history (line) and its 16th--84th percentile range (shaded).}
\label{fig:sed}
\end{figure*}

The fit model reproduces all 20 bands to within $1.7\sigma$ ($\chi^2=13.2$), including the F300M and F356W excesses produced by \hb$+$\oiii and \ha (Fig.~\ref{fig:sed}b), and the redshift posterior agrees with the spectroscopic redshift.
The stellar mass quoted in Section~\ref{sec:data} is the surviving mass; the total mass formed is $0.03$~dex higher.
The star formation history rises towards the present (Fig.~\ref{fig:sed}c): half of the stellar mass formed ($t_{50}$) within the last $7.7^{+10.5}_{-5.0}$~Myr, and the star formation rate averaged over the last 10~Myr exceeds that over the last 100~Myr by a factor of about six.
The diffuse optical depth, $\hat{\tau}_\mathrm{dust,2}=0.04^{+0.03}_{-0.02}$ ($A_V\approx0.04$~mag), is consistent with the observed Balmer decrement.
The stellar metallicity is only weakly constrained, and the attenuation-curve slope and birth-cloud ratio follow their priors.
The photometric gas-phase metallicity, $\log(Z_\mathrm{gas}/Z_\odot)=-1.47^{+0.29}_{-0.33}$ or $\logoh\simeq7.2$, lies above our spectroscopic abundance, but it is constrained only through photometry, and the nebular grid does not extend below $\log(Z_\mathrm{gas}/Z_\odot)=-2$ ($\logoh=6.69$), so we adopt the $R3$-based abundance throughout.

\section{Stellar mass of a young ionizing population from its \texorpdfstring{\ha}{H-alpha} luminosity}
\label{app:halpha_mass}

This appendix gives a general recipe for converting the \ha luminosity of a young star-forming system into the mass of the stellar population that ionizes it, for any assumed IMF, and applies it to \target.

\subsection{Relation between \texorpdfstring{\ha}{H-alpha} luminosity and stellar mass}

For ionization-bounded gas under Case~B recombination, the \ha luminosity is proportional to the rate of hydrogen-ionizing photons absorbed by the gas \citep{Osterbrock2006agna.book.....O,Schaerer2003A&A...397..527S,Lipatova2025arXiv250809331L},
\begin{equation}
L_{\mathrm{H}\alpha}=c_{\mathrm{H}\alpha}\,f_\mathrm{cov}\,Q_\mathrm{H},
\label{eq:lha}
\end{equation}
where $Q_\mathrm{H}$ is the rate at which the stars emit photons above 13.6~eV, $f_\mathrm{cov}=1-f_\mathrm{esc}$ is the fraction of these photons absorbed by the gas, and $c_{\mathrm{H}\alpha}=4\pi j_{\mathrm{H}\alpha}/(n_\mathrm{e}n_\mathrm{p}\alpha_\mathrm{B})$ is the \ha energy emitted per absorbed ionizing photon.
For a coeval population of mass $M_\star$ formed with an IMF $\phi(m)\equiv\mathrm{d}N/\mathrm{d}m$ between $m_\mathrm{l}$ and $m_\mathrm{u}$, the photon rate is $Q_\mathrm{H}=\langle q_\mathrm{H}\rangle M_\star$, with
\begin{equation}
\langle q_\mathrm{H}\rangle=\frac{\int_{m_\mathrm{l}}^{m_\mathrm{u}}q_\mathrm{H}(m)\,\phi(m)\,\mathrm{d}m}{\int_{m_\mathrm{l}}^{m_\mathrm{u}}m\,\phi(m)\,\mathrm{d}m},
\label{eq:qmean}
\end{equation}
where $q_\mathrm{H}(m)$ is the ionizing photon rate of a star of initial mass $m$.
Combining equations~(\ref{eq:lha}) and (\ref{eq:qmean}) gives the stellar mass required by an observed \ha luminosity,
\begin{equation}
\begin{split}
M_\star&=\frac{L_{\mathrm{H}\alpha}}{c_{\mathrm{H}\alpha}\,f_\mathrm{cov}\,\langle q_\mathrm{H}\rangle}\\
&\simeq7.35\times10^{4}\,M_\odot\left(\frac{L_{\mathrm{H}\alpha}}{10^{41}\,\mathrm{erg\,s^{-1}}}\right)f_\mathrm{cov}^{-1}\\
&\quad\times\left(\frac{\langle q_\mathrm{H}\rangle}{10^{48}\,\mathrm{s^{-1}}\,M_\odot^{-1}}\right)^{-1}\left(\frac{c_{\mathrm{H}\alpha}}{1.36\times10^{-12}\,\mathrm{erg}}\right)^{-1},
\end{split}
\label{eq:mass}
\end{equation}
where $L_{\mathrm{H}\alpha}=4\pi D_\mathrm{L}^{2}F_{\mathrm{H}\alpha}\,10^{0.4A_{\mathrm{H}\alpha}}/\mu$ is the intrinsic luminosity for an observed total flux $F_{\mathrm{H}\alpha}$, dust attenuation $A_{\mathrm{H}\alpha}$, and lensing magnification $\mu$.

The normalization $c_{\mathrm{H}\alpha}=1.36\times10^{-12}$~erg is the conventional Case~B value, for $T_\mathrm{e}=10^{4}$~K and $n_\mathrm{e}=10^{2}\,\mathrm{cm^{-3}}$ with an intrinsic $\mathrm{H\alpha/H\beta}=2.86$, computed from the emissivities of \citet{Storey1995MNRAS.272...41S} with \textsc{PyNeb} \citep{Luridiana2015A&A...573A..42L} and the recombination coefficient of \citet{Hui1997MNRAS.292...27H}.
Hotter gas lowers it to $1.27\times10^{-12}$ and $1.21\times10^{-12}$~erg at $T_\mathrm{e}=2\times10^{4}$ and $3\times10^{4}$~K, raising $M_\star$ by 0.03 and 0.05~dex, whereas densities up to $10^{6}\,\mathrm{cm^{-3}}$ change it by less than 1 per cent.
Case~B neglects collisional excitation, which matters in hot, metal-free gas: the photoionization models of the Yggdrasil code \citep{Zackrisson2011ApJ...740...13Z} yield 1.1--1.3 times more \ha per ionizing photon for \piii populations younger than 2~Myr, with $\mathrm{H\alpha/H\beta}\simeq3.1$--3.3, and hence 0.05--0.13~dex lower masses, while their models at $Z=0.0004$ match Case~B to within 2 per cent.
The normalization $\langle q_\mathrm{H}\rangle=10^{48}\,\mathrm{s^{-1}}\,M_\odot^{-1}$ is typical of a top-heavy \piii IMF, whereas canonical IMFs give $\langle q_\mathrm{H}\rangle\simeq(3$--$6)\times10^{46}\,\mathrm{s^{-1}}\,M_\odot^{-1}$ (Table~\ref{tab:qh}), for which the coefficient of equation~(\ref{eq:mass}) becomes $(1.2$--$2.5)\times10^{6}\,M_\odot$.

Equation~(\ref{eq:mass}) assumes a single burst with a fully sampled IMF, and ignores the stochastic sampling that matters for the least massive systems \citep[see][]{Lipatova2025arXiv250809331L,Rusta2025ApJ...989L..32R}.
For zero-metallicity IMFs we adopt lifetime-averaged stellar rates \citep{Schaerer2002A&A...382...28S,Tanikawa2020MNRAS.495.4170T,Deng2024A&A...691A.231D}, which match the instantaneous rates of burst models to within 0.15~dex up to an age of 2~Myr.
The rates then decline, by 0.5--0.7~dex at 5~Myr for the \piii.2 and \piii Kroupa IMFs, and a \piii.1 population has no stars left after 3.6~Myr \citep{Schaerer2002A&A...382...28S,Zackrisson2011ApJ...740...13Z}.
For canonical IMFs, the tabulated 1-Myr rates change by up to $+0.25$~dex by 3~Myr and fall by up to 0.9~dex by 5~Myr, depending on metallicity.
Because $f_\mathrm{cov}\leq1$ and $\langle q_\mathrm{H}\rangle$ is close to its maximum at these ages, equation~(\ref{eq:mass}) returns the minimum mass formed for a given IMF; leakage of ionizing photons, their absorption by dust inside the nebula, older ages, and extended star formation all raise it.

The same route applies to the \heii recombination lines, with $Q_{\mathrm{He^+}}$ the rate of photons above 54.4~eV.
For equal covering of H- and He$^+$-ionizing photons, the line ratio
\begin{equation}
\frac{F_\mathrm{He\,II\,\lambda1640}}{F_{\mathrm{H}\alpha}}=\frac{c_{1640}}{c_{\mathrm{H}\alpha}}\,\frac{\langle q_{\mathrm{He^+}}\rangle}{\langle q_\mathrm{H}\rangle}\simeq4.7\,\frac{\langle q_{\mathrm{He^+}}\rangle}{\langle q_\mathrm{H}\rangle}
\label{eq:heii}
\end{equation}
is independent of $M_\star$, $\mu$, and $f_\mathrm{cov}$, with $c_{1640}=6.35\times10^{-12}$~erg and $c_{4686}=9.8\times10^{-13}$~erg for the same Case~B conditions.
Between $T_\mathrm{e}=10^{4}$ and $3\times10^{4}$~K the ratio $c_{1640}/c_{\mathrm{H}\alpha}$ changes by less than 1 per cent, whereas $c_{4686}/c_{\mathrm{H}\alpha}$ falls by 19 per cent.
If instead all He$^+$-ionizing photons are absorbed while H-ionizing photons escape, the ratio scales as $f_\mathrm{cov}^{-1}$ \citep{Schaerer2002A&A...382...28S}.

\subsection{IMF-averaged ionizing photon rates}

Table~\ref{tab:qh} lists $\langle q_\mathrm{H}\rangle$ for commonly used IMFs.
For zero-metallicity stars we integrate equation~(\ref{eq:qmean}) with the stellar rates of \citet{Deng2024A&A...691A.231D}, summing their three bands above 13.6~eV (their eq.~8 with the $Z\leq10^{-8}\,Z_\odot$ coefficients of their table~B2) for stars above 8~$M_\odot$.
Unlike the population synthesis models of \citet{Schaerer2003A&A...397..527S}, \citet{Raiter2010A&A...523A..64R}, and \citet{Zackrisson2011ApJ...740...13Z}, which are computed for a small set of fixed IMFs, these rates given by \citet{Deng2024A&A...691A.231D} are analytic functions of initial mass, so they can be integrated over any IMF, including the A-SLOTH and NEFERTITI IMFs used below.
Below 54.4~eV these rates are fits to the lifetime-averaged models of \citet{Schaerer2002A&A...382...28S}.
Integrated over the IMFs, they exceed the zero-age population values of \citet{Schaerer2003A&A...397..527S} and \citet{Raiter2010A&A...523A..64R} by 0.13~dex and agree with the Yggdrasil stellar spectra at 1--2~Myr to within 0.1~dex.
The zero-metallicity IMFs are the Yggdrasil \piii.1 (Salpeter slope over 50--500~$M_\odot$), \piii.2 (log-normal in $\ln m$ with characteristic mass 10~$M_\odot$ and dispersion 1 over 1--500~$M_\odot$; \citealt{Tumlinson2006ApJ...641....1T,Raiter2010A&A...523A..64R}), and \piii Kroupa (0.1--100~$M_\odot$) cases, the A-SLOTH best fit ($\mathrm{d}N/\mathrm{d}m\propto m^{-1.77}$ over 13.6--197~$M_\odot$; \citealt{Hartwig2024MNRAS.535..516H}), and the NEFERTITI IMF ($\mathrm{d}N/\mathrm{d}m\propto m^{-2.35}\exp(-10\,M_\odot/m)$ over 0.8--1000~$M_\odot$; \citealt{Larson1998MNRAS.301..569L,Rusta2025ApJ...989L..32R}).
The table also gives the zero-age hardness $\log(\langle q_{\mathrm{He^+}}\rangle/\langle q_\mathrm{H}\rangle)_0$ of the \citet{Schaerer2002A&A...382...28S} stellar models, which sets the maximum \heii strength through equation~(\ref{eq:heii}) and drops by more than 1~dex within 2~Myr.
For canonical IMFs we use FSPS with MIST isochrones \citep{Conroy2009ApJ...699..486C,Conroy2010ApJ...712..833C,Choi2016ApJ...823..102C,Dotter2016ApJS..222....8D}, the stellar population code underlying \textsc{Prospector}, and integrate the spectra of 1-Myr-old populations below 912~\AA\ for the \citet{Salpeter1955ApJ...121..161S}, \citet{Kroupa2001MNRAS.322..231K}, and \citet{Chabrier2003PASP..115..763C} IMFs over 0.08--120~$M_\odot$.
Changing the upper mass limit to 100 or 300~$M_\odot$ shifts these rates by $-0.09$ and $+0.10$~dex, and the Starburst99 \citep{Leitherer1999ApJS..123....3L} and \citet{Schaerer2003A&A...397..527S} models agree with them to within 0.2~dex.

\begin{table}
\centering
\caption{IMF-averaged H ionizing photon rates per unit stellar mass formed, and IMF-averaged He-to-H ionizing rate ratio.
Zero-metallicity rates are lifetime averages, appropriate for ages $\lesssim2$~Myr; canonical-IMF rates are for 1-Myr-old populations at $Z=0.006\,Z_\odot$, with solar-metallicity values in parentheses.}
\label{tab:qh}
\begin{tabular}{lcc}
\toprule
IMF & $\log\langle q_\mathrm{H}\rangle$ & $\log(\langle q_{\mathrm{He^+}}\rangle/\langle q_\mathrm{H}\rangle)_0$ \\
 & [$\mathrm{s^{-1}}\,M_\odot^{-1}$]\\
\midrule
\multicolumn{3}{l}{\textit{Zero metallicity}} \\
\piii.1 & 48.11 & $-0.90$ \\
A-SLOTH & 47.94 & $-1.04$ \\
NEFERTITI & 47.89 & $-0.92$ \\
\piii.2 & 47.68 & $-1.21$ \\
\piii Kroupa & 46.90 & $-1.47$ \\
\midrule
\multicolumn{3}{l}{\textit{Normal metallicity, 0.08--120~$M_\odot$}} \\
Chabrier & 46.78 (46.71) & -- \\
Kroupa & 46.75 (46.69) & -- \\
Salpeter & 46.52 (46.45) & -- \\
\bottomrule
\end{tabular}
\end{table}

\subsection{Application to \texorpdfstring{\target}{LATED-1}}

For \target, the total flux $F_{\mathrm{H}\alpha}=(120.5\pm6.8)\times10^{-20}\,\mathrm{erg\,s^{-1}\,cm^{-2}}$ (Section~\ref{sec:data}), $\mu=2.80$, and no attenuation, as the Balmer decrement of $2.43\pm0.45$ allows, give $L_{\mathrm{H}\alpha}=(1.02\pm0.06)\times10^{41}\,\mathrm{erg\,s^{-1}}$ and $Q_\mathrm{H}=7.5\times10^{52}\,\mathrm{s^{-1}}$ for $f_\mathrm{cov}=1$.
Canonical IMFs then require $\log(M_\star/M_\odot)=6.1$--6.4, consistent with the \textsc{Prospector} mass of $\log(M_\star/M_\odot)=6.30^{+0.19}_{-0.16}$.
Top-heavy \piii IMFs require $\log(M_\star/M_\odot)=4.8$--5.2, that is 3--8 per cent of the \textsc{Prospector} mass, and a \piii population with a Kroupa IMF requires $\log(M_\star/M_\odot)=6.0$.
Adopting $f_\mathrm{cov}=0.5$ raises every value by 0.30~dex and $\mu=1.99$ \citep{Richard2021A&A...646A..83R} by 0.15~dex, whereas the flux uncertainty contributes only 0.03~dex.
Measuring \ha in the instantaneous-burst spectra of the three Yggdrasil \piii models returns masses within 0.02~dex of equation~(\ref{eq:mass}) at 1~Myr.
Normalised to the observed \ha, the same spectra also predict the UV luminosity, which fixes the age of a \piii population for a given covering fraction.
With $f_\mathrm{cov}=1$, the \piii populations younger than 2~Myr give $M_\mathrm{UV}=-15.5$ to $-15.75$, 0.4--0.65~mag fainter than observed, while the \piii.2 and \piii Kroupa populations reach the observed $M_\mathrm{UV}$, or its dust-free value of $-16.40$, at 6--10~Myr with $10^{6.2}$--$10^{6.8}\,M_\odot$.
At 2--3~Myr the \piii.1 and \piii.2 populations reproduce both luminosities with $f_\mathrm{cov}=0.36$--0.51 and $10^{5.1}$--$10^{5.7}\,M_\odot$.
The same spectra predict $F_\mathrm{He\,II\,\lambda1640}/F_{\mathrm{H}\alpha}\leq0.35$ ($F_\mathrm{He\,II\,\lambda4686}/F_{\mathrm{H}\alpha}\leq0.036$) at any age, falling below 0.01 within 2~Myr, against our aperture-corrected $2\sigma$ limits of 0.47 and 0.15 on these ratios.
Case~B with the zero-age hardness of Table~\ref{tab:qh} gives higher values of $F_\mathrm{He\,II\,\lambda1640}/F_{\mathrm{H}\alpha}$, 0.58 and 0.56 for the \piii.1 and NEFERTITI IMFs, which remain below the $3\sigma$ limit of 0.71.
For comparison, our \textsc{Prospector} SED fitting with Pop~II models gives $F_\mathrm{He\,II\,\lambda1640}/F_{\mathrm{H}\alpha}=9_{-3}^{+6}\times10^{-3}$ ($F_\mathrm{He\,II\,\lambda4686}/F_{\mathrm{H}\alpha}=1.4_{-0.5}^{+1.0}\times10^{-3}$).
The \heii limits therefore do not exclude a \piii contribution to the ionizing photon budget of \target.

\section{Comparison samples, \texorpdfstring{$R3$}{R3} calibrations, and photoionization models}
\label{app:r3}

\begin{figure*}
\centering
\includegraphics[width=\textwidth]{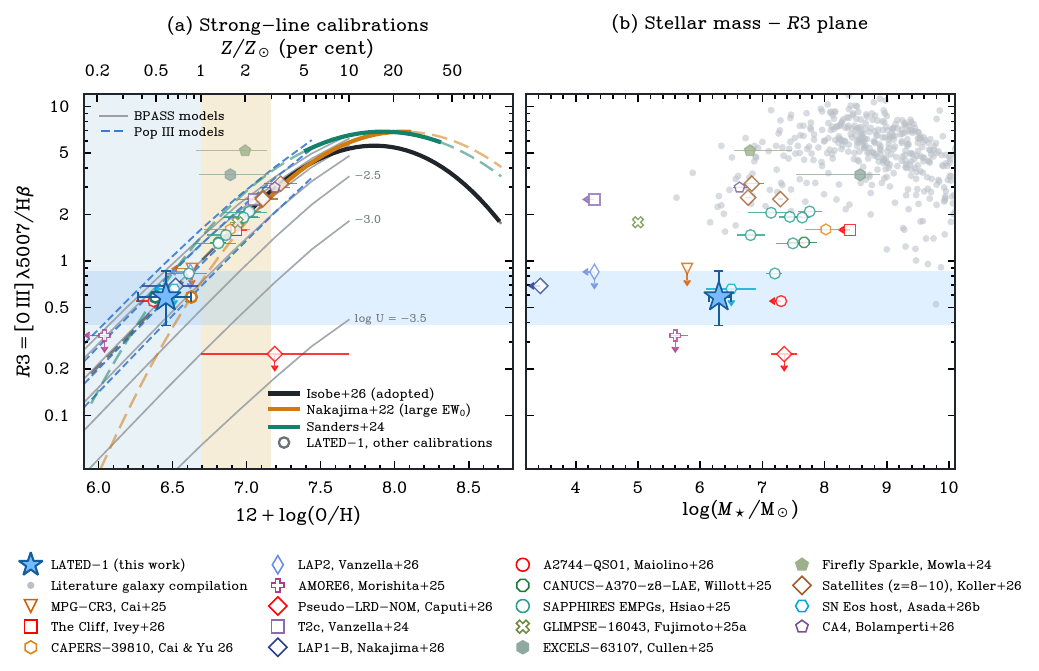}
\caption{$R3$ against oxygen abundance (a) and stellar mass (b) for \target (star) and the comparison systems of Fig.~\ref{fig:models}.
(a) The relations of \citet{Isobe2026arXiv260611345I} (adopted), \citet{Nakajima2022ApJS..262....3N} for galaxies with large $\mathrm{EW_0}(\mathrm{H}\beta)$, and \citet{Sanders2024ApJ...962...24S}. They are denoted by solid lines over their calibrated abundance ranges and dashed where extrapolated. Open circles mark where the measured $R3$ of \target meets the low-metallicity branches of the two alternative relations.
Thin lines are \textsc{Cloudy} models for a 1-Myr BPASS burst (grey, from $\log U=-0.5$ at the top to $-3.5$ at the bottom in steps of 0.5) and for a zero-age \piii population (blue dashed, $\log U=-0.5$ to $-2$), and the vertical bands mark 1--3 per cent (yellow) and less than 1 per cent (blue) of the solar oxygen abundance.
(b) Grey points are the galaxies of Fig.~\ref{fig:models} with a published $R3$ measurement.
In both panels the horizontal band is the 16th--84th percentile range of the $R3$ of \target, the comparison systems keep their symbols and abundances from Fig.~\ref{fig:models}, and arrows mark upper limits on $R3$ ($2\sigma$, except for the pseudo-LRD-NOM, for which no level is quoted) and upper bounds on stellar mass.}
\label{fig:r3cal}
\end{figure*}

The systems shown with coloured symbols in Figs~\ref{fig:models} and \ref{fig:r3cal} are drawn from \citet{Cai2025ApJ...993L..52C,Ivey2026MNRAS.550g1220I,Cai2026RAA....26k5005C,Vanzella2026A&A...705L..12V,Morishita2025arXiv250710521M,Caputi2026ApJ..1007..203C,Vanzella2024A&A...691A.251V,Nakajima2026Natur.653..363N,Vanzella2023A&A...678A.173V,Maiolino2026MNRAS.548f2109M,Willott2025ApJ...988...26W,Ubler2026arXiv260320360U,Hsiao2025arXiv250503873H,Fujimoto2025arXiv251211790F,Isobe2026arXiv260611345I,Asada2026arXiv260714355A,Bolamperti2026A&A...712A.153B,Koller2026MNRAS.551g1206K,Cullen2025MNRAS.540.2176C,Mowla2024Natur.636..332M}, whose heterogeneous diagnostics, calibrations, and confidence levels make them a context sample rather than a homogeneous abundance ranking.
The pale grey points in Fig.~\ref{fig:models} are 2421 literature entries at $z=1.0$--14.2, each giving both the stellar mass and the gas-phase oxygen abundance of a galaxy.
Of these, 1760 come from the catalogue underlying \citet{Isobe2026arXiv260611345I}, provided by the authors, and the others from published tables and machine-readable catalogues \citep{Troncoso2014A&A...563A..58T,Onodera2016ApJ...822...42O,Sanders2020MNRAS.491.1427S,Curti2020MNRAS.492..821C,Wang2020ApJ...900..183W,Henry2021ApJ...919..143H,Wang2022ApJ...926...70W,Li2023ApJ...955L..18L,Nakajima2023ApJS..269...33N,Curti2024A&A...684A..75C,He2024ApJ...960L..13H,Revalski2024ApJ...966..228R,Morishita2024ApJ...971...43M,Chemerynska2024ApJ...976L..15C,Venturi2024A&A...691A..19V,Sarkar2025ApJ...978..136S,Price2025ApJ...982...51P,Chakraborty2025ApJ...985...24C,Scholte2025MNRAS.540.1800S,Nishigaki2025arXiv251212983N,Asada2026arXiv260120045A,Faisst2026ApJ..1004...22F,Hsiao2026arXiv260506770H,Pollock2026A&A...708A.203P,Raptis2026ApJ..1004L..31R,Rowland2026MNRAS.546f2023R,Sanders2026ApJ..1003..228S,Stanton2026MNRAS.547.stag449S}, including six individual galaxies at $z>10$ \citep{Calabro2024ApJ...975..245C,D'Eugenio2024A&A...689A.152D,Hsiao2024ApJ...973...81H,Alvarez-Marquez2025A&A...695A.250A,Helton2026ApJ..1004L..30H,Scholtz2026MNRAS.545f2107S}.
We exclude one-sided limits, samples pre-selected for low metallicity, objects flagged as AGN, and galaxies already shown with coloured symbols.
Galaxies listed more than once, identified by their JADES, CEERS, or ERO numbers, are kept once, with the values of \citet{Isobe2026arXiv260611345I} where available.
Above $z\sim10$, where \oiii$\lambda5007$ leaves the NIRSpec wavelength range, the abundances rest on MIRI spectroscopy, on NIRSpec reductions extended to $5.5\,\mu$m, or on other emission lines and spectral fits.
Because the set mixes direct-$T_\mathrm{e}$ and several strong-line calibrations and adopts stellar masses for different IMFs, it only provides context rather than a homogeneous sample.
The grey points in Fig.~\ref{fig:r3cal}b are the 427 of these galaxies with a published $R3$ measurement, taken from the same publications or from \citet{Wang2024ApJ...967L..42W,Raptis2025ApJ...989L..55R,Faisst2026ApJS..282...19F}.

Figure~\ref{fig:r3cal}a compares the three $R3$ calibrations used in Section~\ref{sec:interpretation}: the adopted relation of \citet{Isobe2026arXiv260611345I}, \citet{Nakajima2022ApJS..262....3N} relation for galaxies with large $\mathrm{EW_0}(\mathrm{H}\beta)$, and that of \citet{Sanders2024ApJ...962...24S}, which are calibrated over $\logoh=7.0$--8.7, 7.1--8.1, and 7.4--8.3, respectively.
At the measured $R3=0.59$ their low-metallicity branches give $\logoh=6.45$, 6.63, and 6.38, so all three values are extrapolations.
Below $\logoh=7.0$ the adopted relation follows the slope of photoionization models rather than an extended polynomial, and it places \target on the same scale as most of the comparison systems, whose abundances were derived with it; those systems therefore lie on its curve by construction and do not test it.
The two direct-$T_\mathrm{e}$ measurements in the panel, EXCELS-63107 and the Firefly Sparkle at $\logoh\simeq6.9$--7.0 \citep{Cullen2025MNRAS.540.2176C,Mowla2024Natur.636..332M}, lie 0.2--0.5~dex above all three relations in $\log R3$, so individual galaxies can depart from these calibrations by as much as the spread among them or more.

The panel also shows photoionization models computed with \textsc{Cloudy} version~22.00 \citep{Ferland2017RMxAA..53..385F,Chatzikos2023RMxAA..59..327C} in the configuration that \citet{Isobe2026arXiv260611345I} use below $\logoh=7.0$: a 1-Myr instantaneous burst from BPASS v2.2 \citep{Stanway2018MNRAS.479...75S} with a \citet{Salpeter1955ApJ...121..161S} IMF up to $100\,M_\odot$ and stellar metallicity tied to the gas metallicity, a hydrogen density of $300\,\mathrm{cm^{-3}}$, solar abundance ratios, and $-3.5\leq\log U\leq-0.5$.
As in the \citet{Nakajima2022MNRAS.513.5134N} models used for LAP1-B \citep{Nakajima2026Natur.653..363N}, we add a zero-age \piii population, using the Yggdrasil stellar spectrum for a Kroupa IMF over 0.1--100~$M_\odot$ \citep{Zackrisson2011ApJ...740...13Z}, which above $1\,M_\odot$ is close to their Salpeter IMF.
At the measured $R3$, the BPASS models give $\logoh=6.26$, 6.44, 6.60, 6.78, and 7.09 for $\log U=-0.5$, $-1.5$, $-2.0$, $-2.5$, and $-3.0$, and the \piii models agree with them to within 0.05~dex where both are computed.
The adopted calibration therefore corresponds to $\log U\simeq-1.5$, within the range $-2\lesssim\log U\lesssim-1.5$ that reproduces most metal-poor galaxies \citep{Isobe2026arXiv260611345I}, and the abundance stays below one per cent solar for $\log U\gtrsim-2.3$.

Figure~\ref{fig:r3cal}b removes the calibration by plotting the measured $R3$ against stellar mass.
The 26 comparison galaxies with $10^{5.5}<M_\star/M_\odot<10^{7}$ all have $R3\geq1.55$, 2.6 times the value for \target, and only one of the 427 galaxies, at $M_\star\approx10^{9.8}\,M_\odot$, lies below its 84th-percentile value of 0.86.
Above $10^{5.5}\,M_\odot$, comparably low ratios have otherwise been reported only for SAPPHIRES-7695 \citep{Hsiao2025arXiv250503873H} and, as $2\sigma$ upper limits, for AMORE6, MPG-CR3, and the host of SN~Eos \citep{Morishita2025arXiv250710521M,Cai2025ApJ...993L..52C,Asada2026arXiv260714355A}.
The pseudo-LRD-NOM and A2744-QSO1 also have low ratios, but the gas density of the former exceeds the critical density of \oiii$\lambda5007$ \citep{Caputi2026ApJ..1007..203C} and the stellar mass of the latter is only an upper limit \citep{Maiolino2026MNRAS.548f2109M}.
The separation of \target from galaxies of similar mass is therefore present in the measured line ratio itself, independent of the abundance calibration, although a low $R3$ can also arise from a low ionization parameter, as panel~(a) shows.

\label{lastpage}
\end{document}

%% file: author.tex
\ExplSyntaxOn
\prop_new:N \g__affil_text_prop
\prop_new:N \g__affil_num_prop
\seq_new:N  \g__affil_used_seq
\int_new:N  \g__affil_count_int
\seq_new:N  \l__affil_keys_seq
\seq_new:N  \l__affil_out_seq
\cs_generate_variant:Nn \prop_gput:Nnn { Nne }
\cs_generate_variant:Nn \seq_put_right:Nn { Ne }
\msg_new:nnn { autoaffil } { undefined }
  { Affiliation~key~'#1'~is~used~but~was~never~declared~with~\iow_char:N\\defaffil. }
\NewDocumentCommand \defaffil { m +m }
  { \prop_gput:Nnn \g__affil_text_prop {#1} {#2} }
\NewDocumentCommand \affil { m }
  {
    \seq_set_from_clist:Nn \l__affil_keys_seq {#1}
    \seq_clear:N \l__affil_out_seq
    \seq_map_inline:Nn \l__affil_keys_seq
      {
        \prop_if_in:NnF \g__affil_text_prop {##1}
          { \msg_error:nnn { autoaffil } { undefined } {##1} }
        \prop_if_in:NnF \g__affil_num_prop {##1}
          {
            \int_gincr:N \g__affil_count_int
            \prop_gput:Nne \g__affil_num_prop {##1} { \int_use:N \g__affil_count_int }
            \seq_gput_right:Nn \g__affil_used_seq {##1}
          }
        \seq_put_right:Ne \l__affil_out_seq
          { \exp_not:N \hyperlink { affil:##1 } { \prop_item:Nn \g__affil_num_prop {##1} } }
      }
    \textsuperscript { \seq_use:Nn \l__affil_out_seq { , } }
  }
\NewDocumentCommand \printaffils { }
  {
    \seq_map_inline:Nn \g__affil_used_seq
      {
        \hypertarget { affil:##1 } { \textsuperscript { \prop_item:Nn \g__affil_num_prop {##1} } }
        \prop_item:Nn \g__affil_text_prop {##1} \par
      }
  }
\ExplSyntaxOff

\RequirePackage{orcidlink}
\newcommand{\orcidsymb}[2]{\mbox{#1$^{\mbox{\orcidlink{#2}}}$}}

\newcommand{\corrmark}{$^{\star}$}
\newcommand{\corremail}[1]{\par\corrmark E-mail: \href{mailto:#1}{#1}}